\pdfoutput=1
\documentclass[cits]{JINST}

\usepackage{url}
\usepackage{wasysym}
\newcommand {\beq}       {\begin{equation}}
\newcommand {\eeq}       {\end{equation}}

\newcommand {\defuni}[1] {\ifmmode \mathrm{#1} \else $\mathrm{#1}$ \fi}
\newcommand {\um}[1]     {\defuni{\; #1}}

\title{
Analysis and correction of the magnetic field effects in the Hybrid
Photo-Detectors of the RICH2 Ring Imaging Cherenkov detector of LHCb.
}

\author{R. Cardinale$^a$\thanks{Corresponding
author.}~, C. D'Ambrosio$^b$, R. Forty$^b$, C. Frei$^b$, T. Gys$^b$,
A. Petrolini$^a$, D. Piedigrossi$^b$, B. Storaci$^{b,c,}$\thanks{Now at
  NIKHEF National Institute for Subatomic Physics, Amsterdam,
  Netherlands} $\phantom{ }$ and M. Villa$^{b,c,}$\thanks{{Now at Physikalisches
    Institut der Universitat Bonn, Bonn, Germany and European Organization for Nuclear Research (CERN), Geneva, Switzerland}}\\
\llap{$^a$}Dipartimento di Fisica dell'Universit\`a di Genova e INFN sezione di
Genova,\\
  Genova, Italy\\
\llap{$^b$}European Organization for Nuclear Research (CERN),\\
  Geneva, Switzerland\\
\llap{$^c$}Dipartimento di Fisica dell'Universit\`a di Milano Bicocca e INFN
sezione di Milano Bicocca,\\ Milano, Italia.\\
  E-mail: \email{roberta.cardinale@ge.infn.it}}

\abstract{
The Ring Imaging Cherenkov detectors of the LHCb experiment at the Large Hadron
Collider at CERN are equipped with Hybrid Photo-Detectors. These vacuum
photo-detectors are affected by the stray magnetic field of the LHCb magnet,
which degrades their imaging
properties. This effect increases
the error on the Cherenkov angle measurement and would reduce the particle identification capabilities of LHCb.
A system has been developed for the RICH2 Ring
Imaging Cherenkov detector to perform a detailed characterisation of the magnetic
distortion effects. It is described, along with the methods implemented to correct for these effects,
restoring the optimal resolution.
}

\keywords{Cherenkov detectors, Photon Detectors for UV, visible and IR
photons (vacuum), Detector alignment and calibration methods}
\begin{document}











\section{Introduction}
\label{sec:intro}

\subsection{The LHCb experiment at the LHC}
\label{sec:LHCb}

\begin{figure}[ht]
\centering
\includegraphics[scale=0.25]{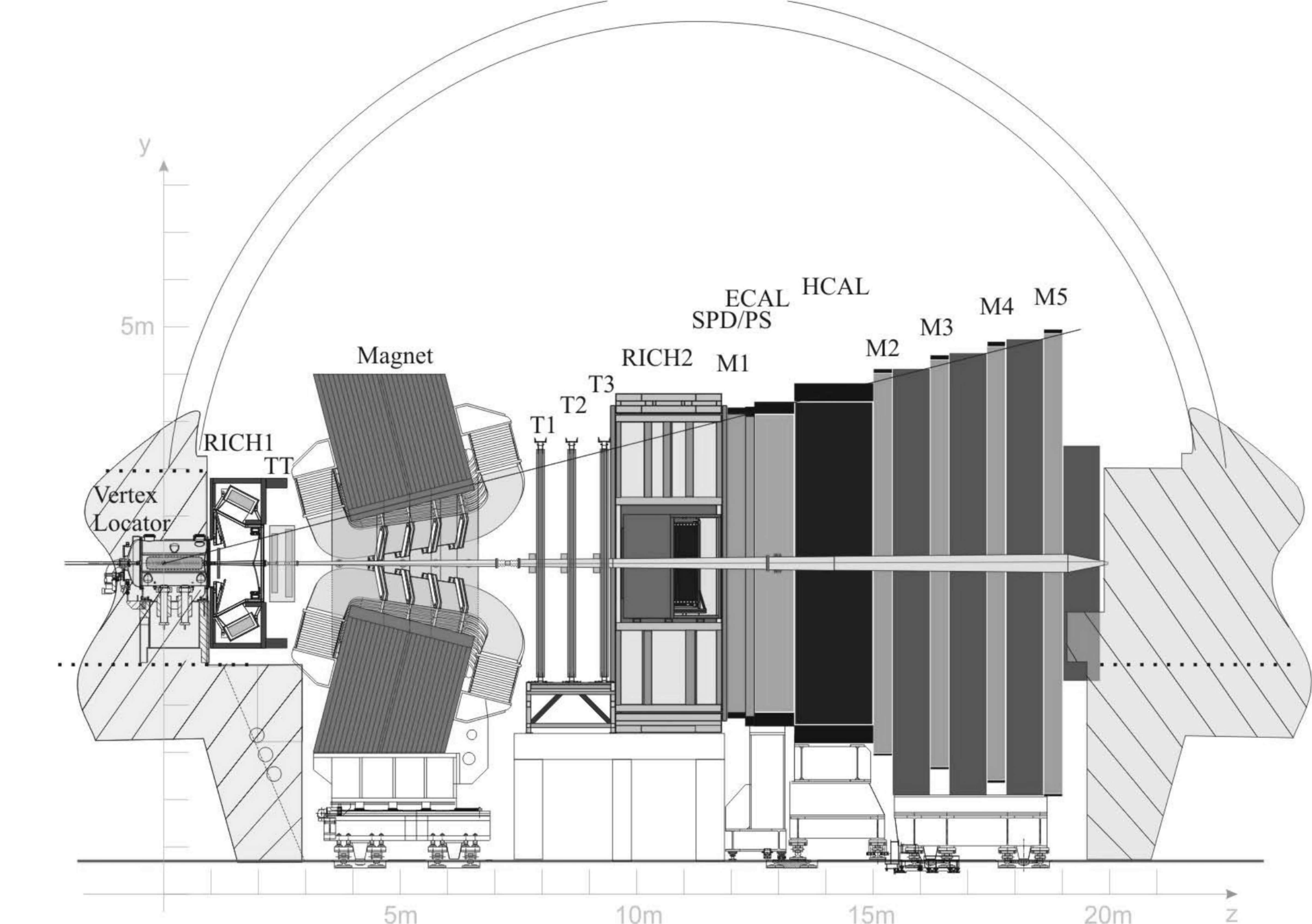} 
\caption{A schematic view of the LHCb detector and its subdetectors.}
\label{fig:LHCb}
\end{figure}

The LHCb experiment~\cite{bi:LHCb} at the Large Hadron Collider (LHC)~\cite{bi:LHC} at CERN
(Geneva, Switzerland) is designed for high precision measurements of CP violation and rare
decays of heavy flavours. Key physics measurements to be performed with LHCb are discussed
in~\cite{bi:RoadMap}. A drawing of the LHCb detector
  is shown in Figure~\ref{fig:LHCb}.

Particle IDentification (PID) of hadrons is an essential tool for the
LHCb physics program. Many final states of interesting decay channels
have background from decays with the same topology, but different
final state hadrons.

\subsection{The RICH detectors of LHCb}
\label{sec:RICH}

The PID capabilities of the LHCb apparatus rely on its two
Ring Imaging Cherenkov detectors, RICH1 and RICH2~\cite{bi:LHCb, bi:CDAmbrosio}.

A single RICH detector cannot satisfy the broad momentum range
requirement. However the strong polar angle and momentum correlation
can be exploited by a system of two RICHes: RICH1 and RICH2. The two
combined sub-detectors can distinguish kaons, pions and protons
in the momentum range $1 \div 100 \um{GeV}/c$ over the full angular
acceptance of LHCb by using three different radiator materials.

The RICH1 detector is located close to the interaction point, upstream
of the LHCb magnet, with $25 - 300\um{mrad}$ polar angular
acceptance. It employs C$_{4}$F$_{10}$ gas and silica aerogel as Cherenkov
radiators providing particle identification of low momentum tracks,
over the range $1-40\um{GeV}/c$.

The RICH2 detector is located approximately half-way between the LHCb magnet and the
large ferromagnetic mass of the muon detector system at a distance along
the beam axis of about $6\um{m}$ from the magnet centre and of about $10\um{m}$ from
the collision point. It is intended for high momentum, up to $100\um{GeV}/c$, and small polar
angle tracks, and uses a gaseous CF$_{4}$ radiator. It covers $10 -
120 \um{mrad}$ in the horizontal plane and $10-100 \um{mrad}$ in the
vertical plane.

Both RICH detectors have a similar optical geometry. Cherenkov
photons,
produced by charged particles traversing the radiator media
above the momentum threshold for Cherenkov light production,
are focused by spherical mirrors and reflected off
flat mirrors onto a photo-sensitive matrix of Hybrid Photo-Detectors
(HPD)~\cite{ bi:LHCb, bi:HPD},
described in Section~\ref{sec:HPD}. Each RICH detector has two HPD planes,
above and below the beam pipe for RICH1, on both sides of the beam
pipe for RICH2.

\begin{figure}[ht]
\centering
\includegraphics[scale=0.5]{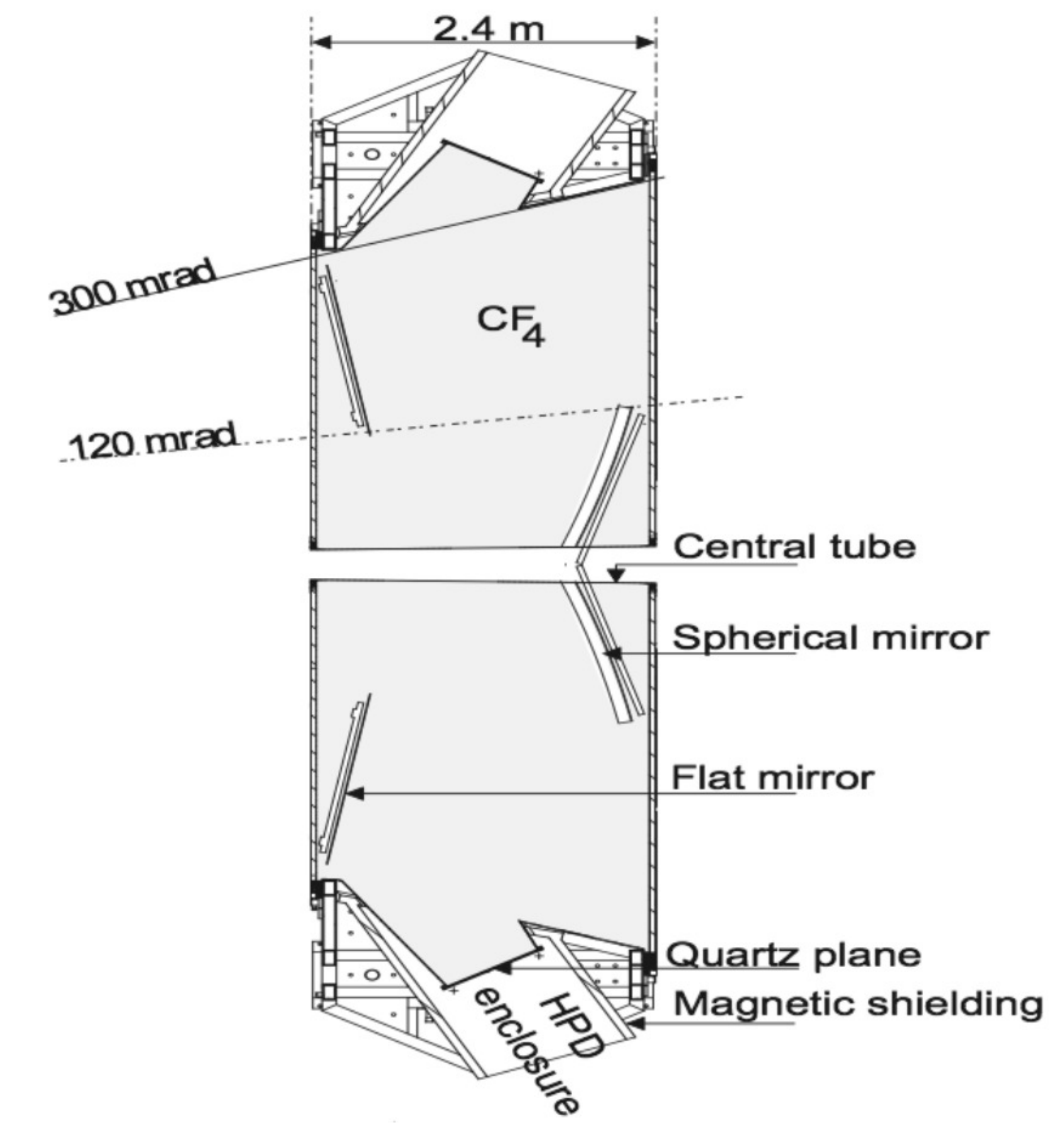} 
\caption{A drawing of RICH2 showing the beam pipe, the spherical
mirrors, the flat mirrors and the HPD enclosure.}
\label{fig:RICH2}
\end{figure}

The RICH detectors are located in the stray field of the LHCb dipole
magnet that has an integrated magnetic field of $4\um{Tm}$. The HPDs
are affected by the stray magnetic field. 
As the HPDs must operate with optimal efficiency and imaging performance, it
is important to understand, measure and correct for its
effects~\cite{bi:Thesis}.
In this paper the strategy to measure and correct the magnetic
distortions in RICH2 are described. Another dedicated system has been
developed for RICH1~\cite{bi:RICH1}.
A drawing of RICH2 is shown in Figure~\ref{fig:RICH2}.

\subsection{The HPD of the RICH detectors}
\label{sec:HPD}

\begin{figure}[htbp]
\centering
\includegraphics[scale=1.1]{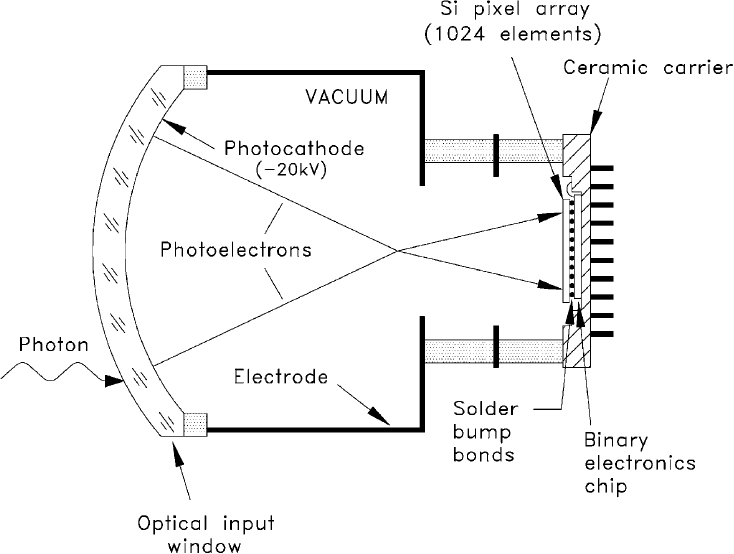}
\caption{ 
A schematic view of the LHCb RICH HPD.
}
\label{fig:HPD}
\end{figure}

A schematic view of an HPD~\cite{bi:HPD} is shown in
Figure~\ref{fig:HPD}.

Photons reaching the HPD photocathode, on the inner-side of the quartz entrance window, produce
photoelectrons which are accelerated and focused by the internal electrostatic
field inside the vacuum tube.
Photoelectrons are detected by a silicon pixel anode chip with  $32$
columns $\times$ $32$ rows of pixels, bump-bonded to a
binary electronics readout chip.
The photocathode is biased at $-18 \um{kV}$ with
respect to the anode.   

The demagnification produced by the electrostatic field is such
that the effective pixel size of $ 0.5 \times 0.5
\um{mm^2} $ on the anode chip corresponds to $ 2.5
\times 2.5 \um{mm^2} $ on the HPD entrance window.

In presence of magnetic field photoelectrons inside the HPD are deviated by the
Lorentz force. 
The imaging properties of the HPD are therefore worsened. If no correction is applied, the error on the Cherenkov angle 
measurement increases, degrading the PID capability of LHCb.

\subsection{The stray magnetic field in RICH2}
\label{sec:MagnField}

The effect of the stray magnetic field on the photoelectrons
trajectories depends on the intensity and direction of the magnetic field
inside the HPD~\cite{bi:BFieldEffect}. In a first approximation, when the
direction of the magnetic field is perpendicular to the geometrical axis of the HPD, there is
an image displacement; on the other hand when the magnetic field is
parallel to this axis,
photoelectrons move along a spiral trajectory causing a rotation of
the image~\cite{bi:BFieldEffect}.

The stray magnetic field inside RICH2 has been calculated~\cite{bi:magnfield} and it
reaches up to $\sim 15\um{mT}$ in amplitude with varying
directions. In order to reduce its effects on the HPDs, the phototubes matrix has been
installed inside a $6\um{cm}$ thick iron shielding box. Moreover each single HPD is
surrounded by a $0.8\um{mm}$ thick cylindrical shield made of Mu-Metal$^{\copyright}$. This shielding effectively reduces the
stray magnetic field inside the HPD. Simulations~\cite{bi:TGnote}
estimate this residual field to
be less than $\approx 1\um{mT}$ with a direction approximately parallel to the
HPD geometrical axis and an absolute value dependent on the HPD position within the
shielding box.

Measurements have been performed showing that if the field inside the HPD volume
is less than $\approx 1\um{mT}$ there is no loss of photoelectrons~\cite{AglieriRinella:2005ef}.

However this residual stray magnetic field produces a
non-negligible image distortion.

A detailed characterisation procedure of the magnetic distortion effects is
necessary to implement a correction. Measurements with magnetic field
on and off have
been performed, in order to parametrise the effects and to determine a correction
procedure restoring the optimal resolution.

\section{Experimental setup}

The correction of the magnetic distortion effect for RICH2 is based on
the projection of a known light pattern onto the two $ 16 \times 9$ HPD matrices,
using a commercial light projector. The pattern is a suitable grid of light spots and its
layout can be chosen to optimise the measurements results.
 
Data with the LHCb magnetic field on and off have been taken and
analysed to compare the position of the light spots in the two conditions.

The ultimate goal is to achieve an accuracy in the correction procedure such
that the residual uncertainty due to magnetic distortion is negligible, in
comparison to the HPD pixel size error~\cite{bi:Thesis}.

\subsection{The light projector}

A commercial light projector~\footnote{Samsung SP-P310ME}~\cite{bi:beamer}
has been used to project the image onto the HPD matrix. 

Its most relevant specifications are summarised in Table~\ref{ta:specs}. 

\begin{table}[htbp]
\centering
\caption{Main specifications of the Samsung SP-P310ME light projector.}
\begin{tabular}{|c|c||}
\hline
Resolution & $800 \times 600 \um{pixel}$ \\
Contrast & $1000:1$\\
Luminosity & 50 ANSI Lumen (1300 Lux) \\
Image size at $2 \um{m}$ distance&
$ 101.6 \times 76.2 \um{cm^2}$
\\
Weight & $0.7\um{Kg}$\\
Dimensions (w $\times$ h $\times$ d) & $12.7\um{cm} \times 5.1\um{cm} \times 9.4\um{cm}$\\
\hline
\end{tabular}
\label{ta:specs}
\end{table}

Two such light projectors have been temporarily installed during the LHCb
commissioning phase inside the
RICH2 gas vessel to shine onto the entire HPD matrix. They were centrally
and symmetrically positioned for the two sides, at mid-height and at a
distance of $\sim 3.3 \um{m}$ from the HPD plane. Neutral filters
have been used in order to reduce the number of photons hitting the
HPDs, to avoid possible damage of the photocathodes.

The choice of using a light projector, instead of a dedicated light source
system, has some advantages. The light projector, thanks to its reduced size, is very
easy to handle, transport and install inside RICH2. 
It does not contribute to the LHCb
detector material budget and it is not necessary for it to be radiation hard.
Moreover it allows to choose the shape of the projected light pattern with a high
spatial resolution.

\subsection{The projected light pattern}

A grid of light spots on a black background has been projected onto
the HPD matrix. In
order to have the best
discrimination for image movements, the smallest possible light spot
size was chosen, corresponding to one pixel of the light projector. Since the light projector has been installed at $\sim 3.3 \um{m}$ distance from the HPD matrix, 
the projected size of the pixel is $2 \times 2 \um{ mm^{2}}$ on the photocathode. This size
is approximately equal to the RICH2 pixel size and it matches well the optical
resolution of the light projector.

\begin{figure}[htbp]
\centering
\includegraphics[scale=0.45]{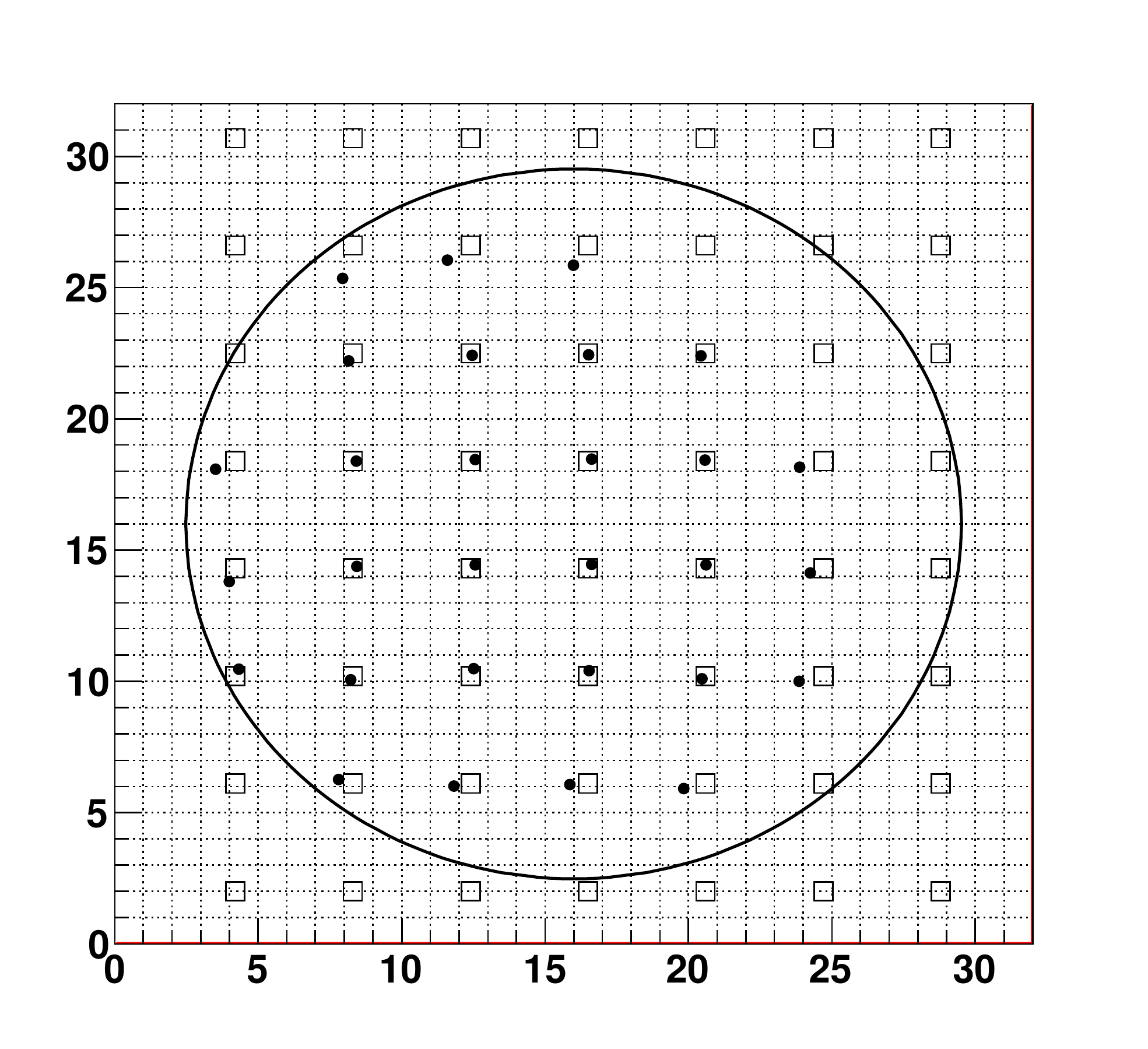}
\caption{ 
One typical HPD chip ($32 \times 32$ pixels matrix) with the projected pattern
(grid of squares) and the reconstructed spots (superimposed dots)
affected by refraction effects and reconstruction errors. Points on
the HPD right side are not reconstructed because the HPD shielding
partially shades the projector light which is not perpendicular with
respect to the HPD plane.}
\label{fig:pattern}
\end{figure}

The grid is made of single bright pixels separated by five dark pixels
(see Figure~\ref{fig:pattern}). This is
a good compromise to have a large enough number of projected light spots, for better statistics, and
the need of a suitable minimum separation between two nearby light spots, 
to avoid superposition of different light spots during the image reconstruction.

The pattern has also three larger spots corresponding to the position of three
dedicated PMTs that have been installed in each of the HPD matrix to be used for aligning
the pattern.

\subsection{The pattern alignment procedure}

Six Hamamatsu H3164-10 PMTs~\cite{bi:pmt} have been installed inside the HPD
matrix, three per side.

These PMTs are used to align the pattern and to know
the exact position of the light spots in the LHCb coordinate frame, since
the PMT position coordinates are precisely known from the survey of
RICH2~\cite{bi:edmsChristoph}. This allows to check that the pattern
position is always the same over time.

In order to align the pattern, the image is moved until a maximum in the signal
coming from the three alignment PMTs is found.
In this way one can
reconstruct the pattern position in the global coordinate system and the
global position of each HPD is known.

The data acquisition system for the alignment PMTs is described in~\cite{bi:TesiMarco}.

\section{Data Analysis}
\label{sec:analysis}

An algorithm to reconstruct the position of the light spots
has been implemented. It can be used both to reconstruct the pattern with and
without magnetic field and for internal alignment purpose of the RICH2 sub-components.

Avoiding a large
amount of light onto the HPD in a single shot requires to analyse
accumulated pictures integrating on many events. The average
number of hits per event has thus been set to be less than one photoelectron per HPD.

\subsection{Reconstruction Algorithm}
In this section the reconstruction algorithm~\cite{bi:Thesis,  bi:barbara} is
briefly summarised.
The algorithm is made of three steps:
\paragraph{Search for local maxima}
The entire HPD chip is scanned, pixel
  by pixel,
  using a recursive function that looks for the absolute maximum inside a
  subregion surrounding 
  the pixel. Whenever a candidate maximum is found, the search region is
  recentered around the candidate maximum and the search for the
  maximum is repeated. If a different pixel is found with a bigger number of hits,
  it becomes the new candidate maximum, the previous one is
  rejected and the procedure is repeated again.\\
  The absolute maximum found in the subregion is rejected if its number of
  hits is smaller than the median value computed on the entire HPD
  matrix.\\ All the pixels labelled as candidate maxima are then
  excluded and the search continues in the rest of the HPD chip. 
\paragraph{Clustering}
A recursive function looks for other local maxima
  in the region
  around the absolute maxima found at the previous step and builds the cluster.
\paragraph{Estimation of the cluster center}
Two different methods have been
  implemented and compared: the average of the position of the pixels
  in the cluster weighted by the number of hits and a
  Gaussian fit of the cluster profile. They give compatible
  results. In both cases the standard deviation of the reconstructed
  centroids is $\lesssim 0.45\um{pixel}$.

Two main factors can deteriorate the reconstruction procedure: noisy pixels and ion-feedback
  effects: they are described below.

\subsubsection{Noisy pixel masking}
\label{subsub:noisy} 

During the analysis procedure noisy pixels ($< 1\permil$ of the pixels) in the HPDs have been identified using
the reconstruction algorithm so that a masking procedure has been implemented.  
In fact it is important to apply a procedure that can distinguish between noisy pixels
and genuine local light maxima, because a noisy pixel can satisfy the
conditions to be considered as a local maximum. 
All real local light maxima, that is local maxima corresponding to a light
spot, have approximately the same number of hits,
since the projected light is uniform. On the contrary noisy pixels have a number of counts
that is usually larger by an order of magnitude. Therefore all local
maxima with a factor of two more counts with
respect to other local maxima found in the same HPD, are rejected.  
The different parameters were tuned so that 
noisy pixels are rejected without losing any real local light maximum.

\subsubsection{Ion-feedback reduction}
\label{subsub:ifb} 

Residual gas inside the HPD can cause
ion-feedback~\cite{bi:IonFeedBack}; consisting in a
photoelectron which ionises molecules of the residual gas, creating positive
ions which then drift to the photocathode. Ions hitting the photocathode
produce secondary electron emission. A ion-feedback event is characterised by a large number of hits grouped
in clusters in the HPD centre.

For the majority of the HPDs this effect only gives a small noise increase. However for $10\%-20\%$ of the HPDs the ion-feedback rate
can be so high that it spoils the signal and
provokes accelerated ageing of the photocathode and a decrease of the detection efficiency.

If the ion-feedback background is large, the algorithm may have
problems to reconstruct the light spot positions. 

\begin{figure}[ht]
\centering
\begin{tabular}{cc}

\includegraphics[scale=0.37]{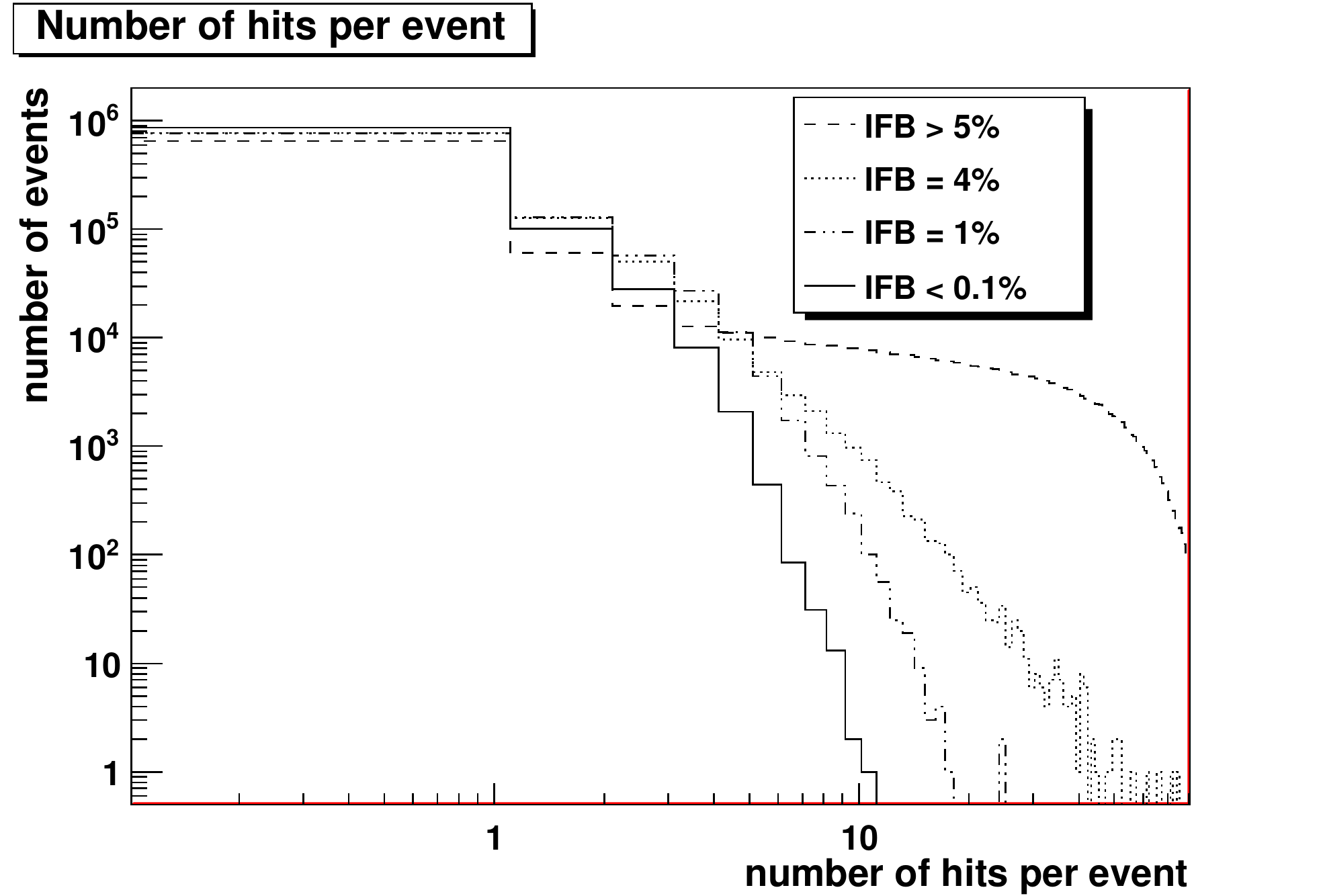}
&
\includegraphics[scale=0.37]{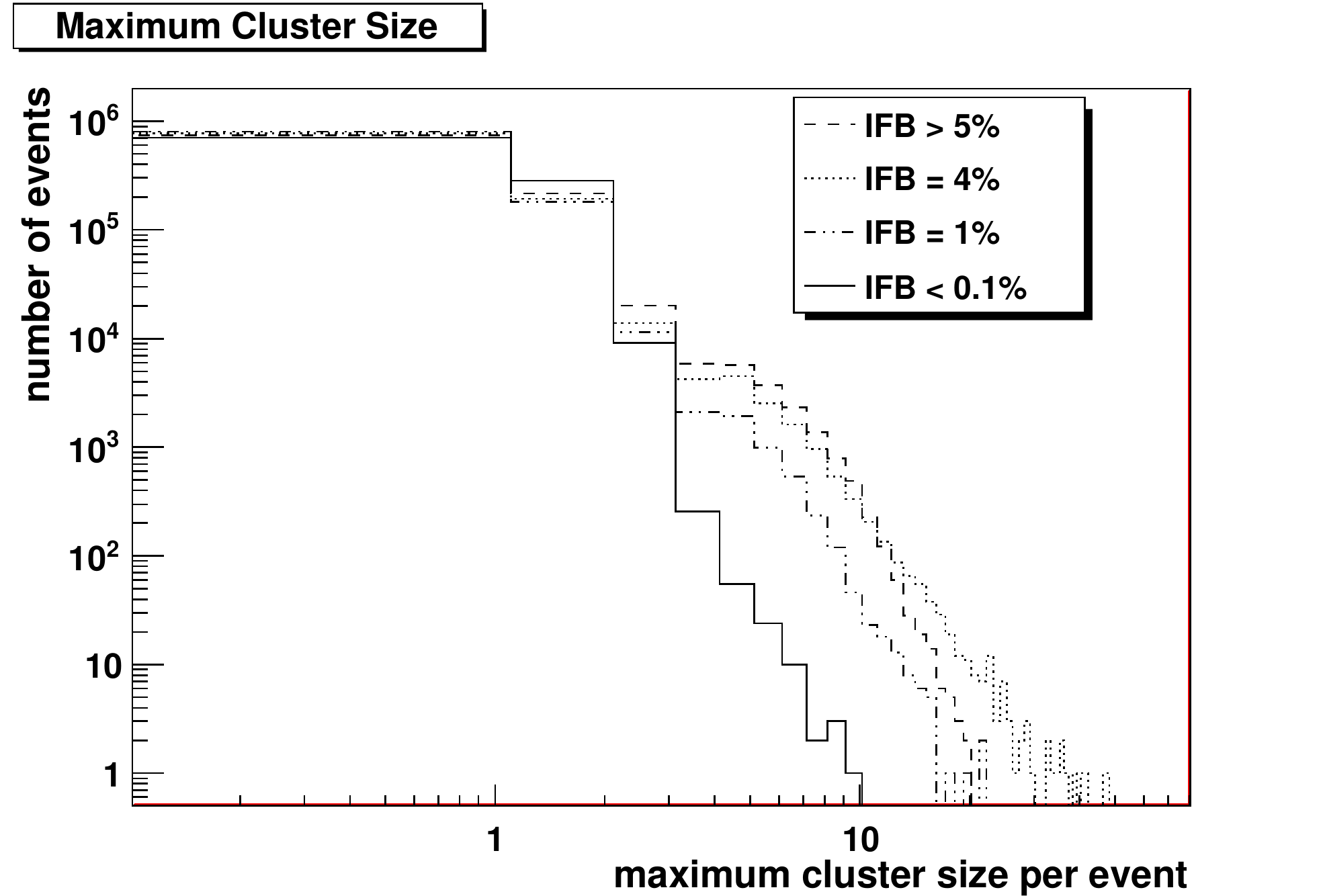}
\\
\includegraphics[scale=0.37]{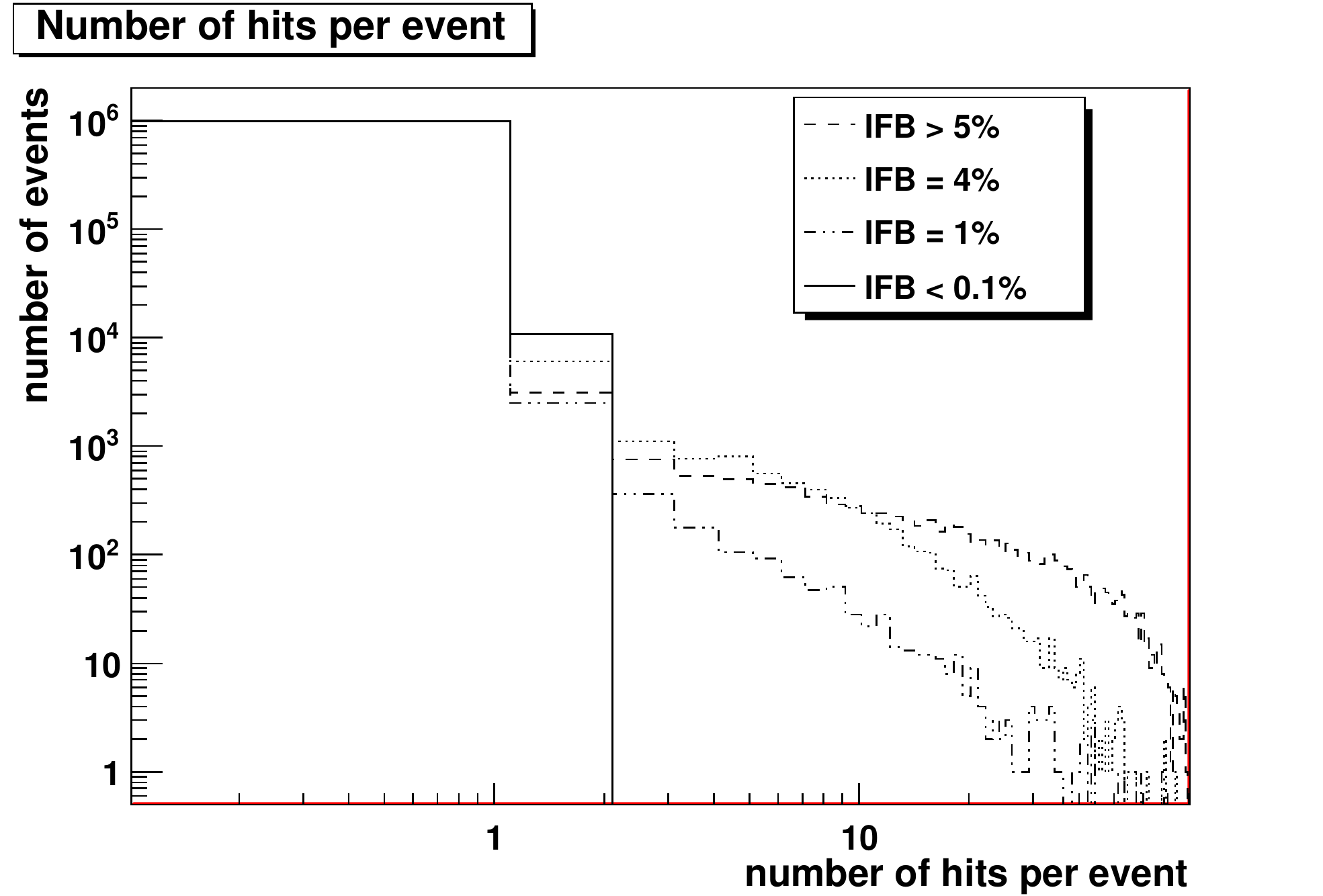}
&
\includegraphics[scale=0.37]{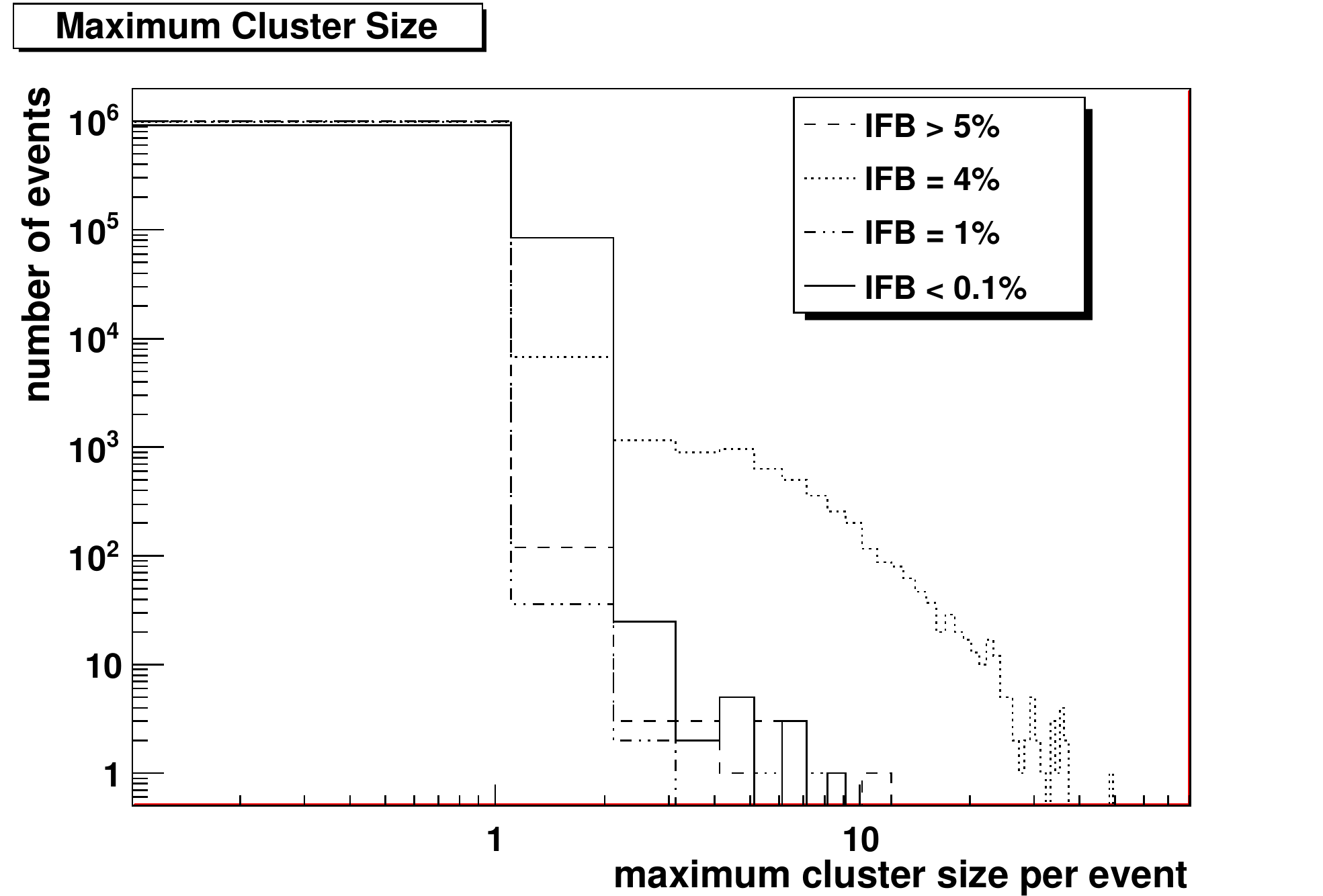}\\
\\
\end{tabular}
\caption{
Histogram of the number of hits and maximum cluster size, used to
characterise the ion-feedback effect in data with the light pattern (top) and
dark count data (bottom).
}
\label{fig:ifb}
\end{figure}

\begin{figure}[htb]
\centering
\begin{tabular}{cc}
\includegraphics [scale=0.35] {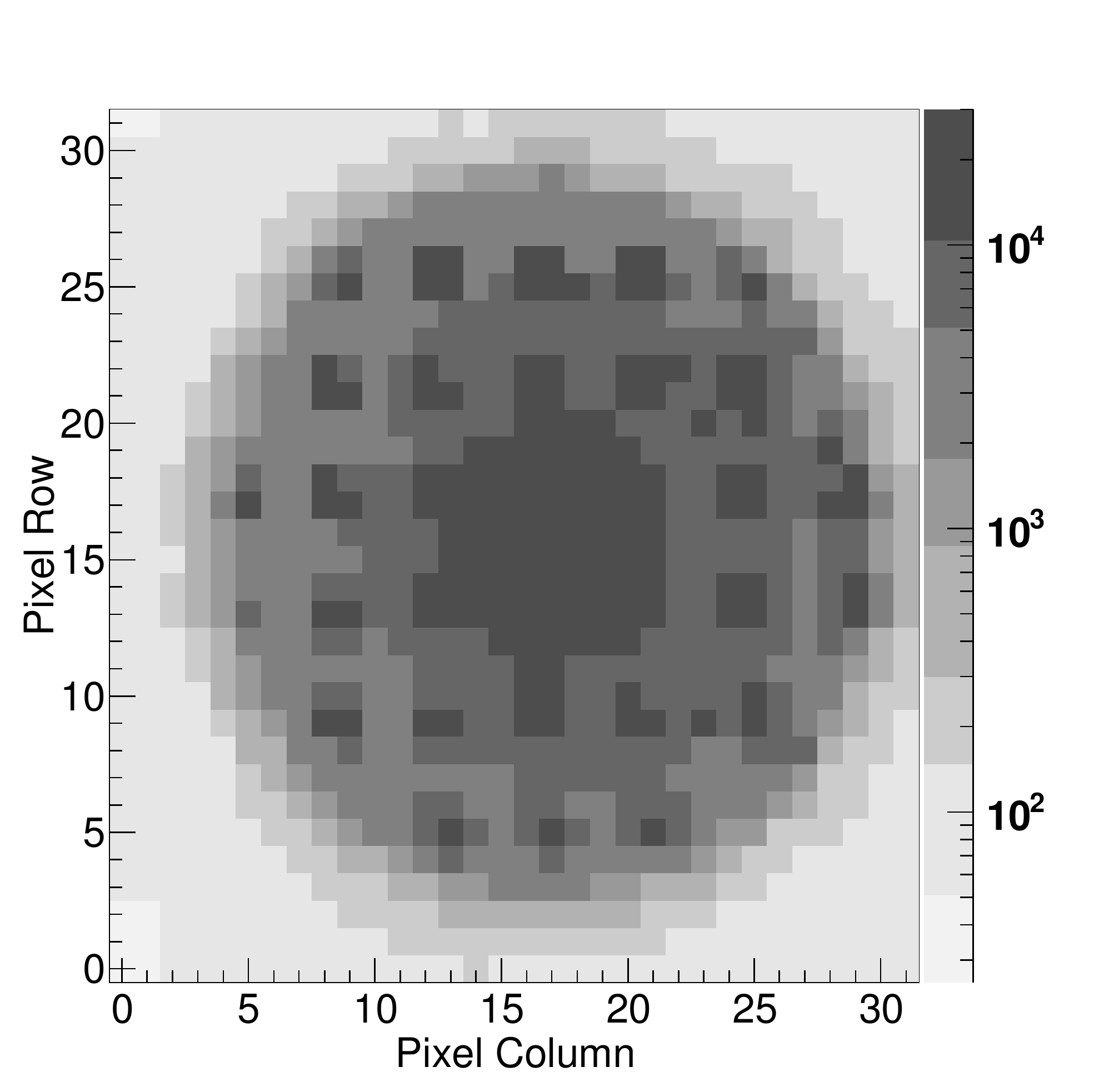} &
\includegraphics [scale=0.35] {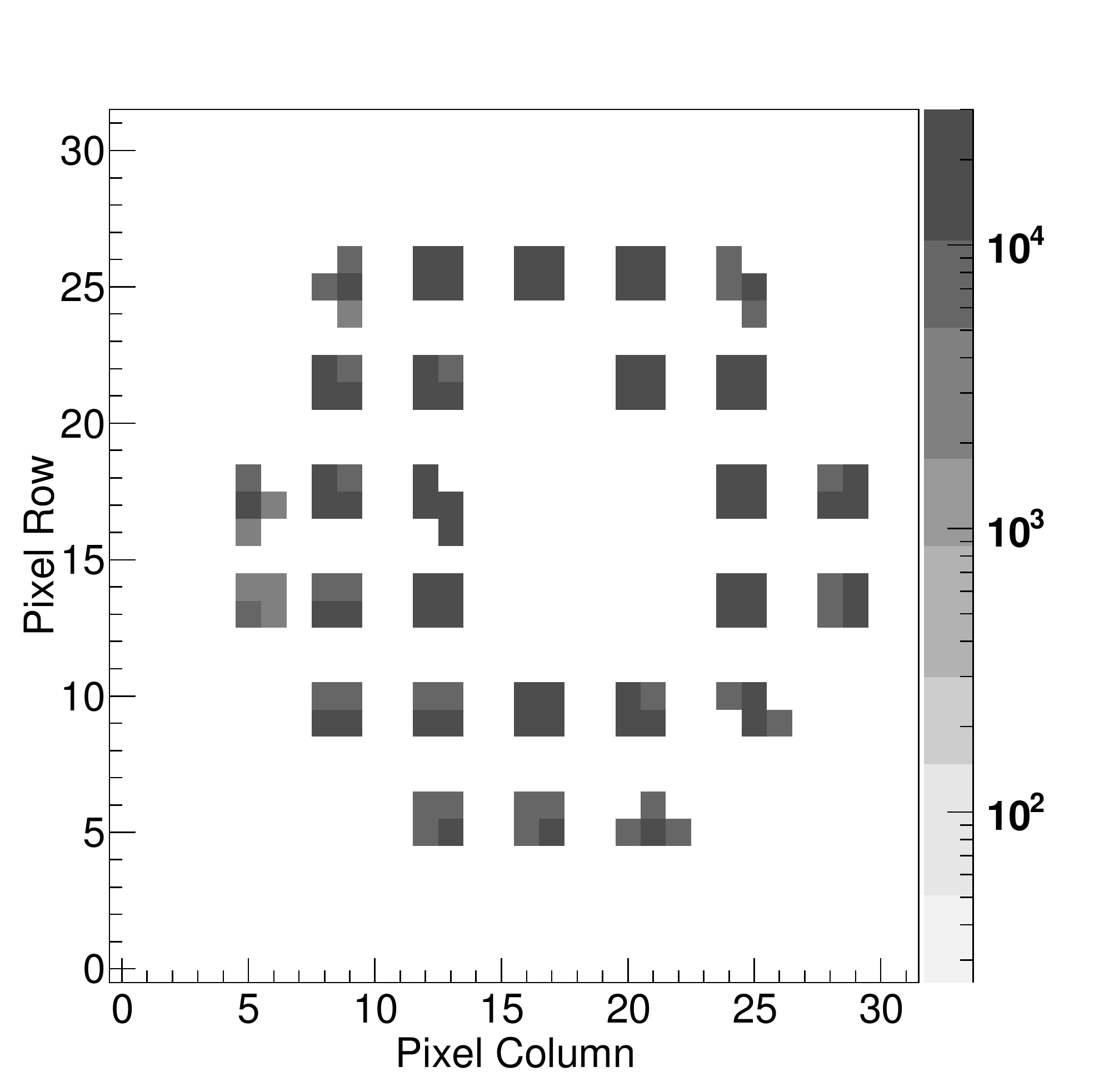} \\
\includegraphics [scale=0.35] {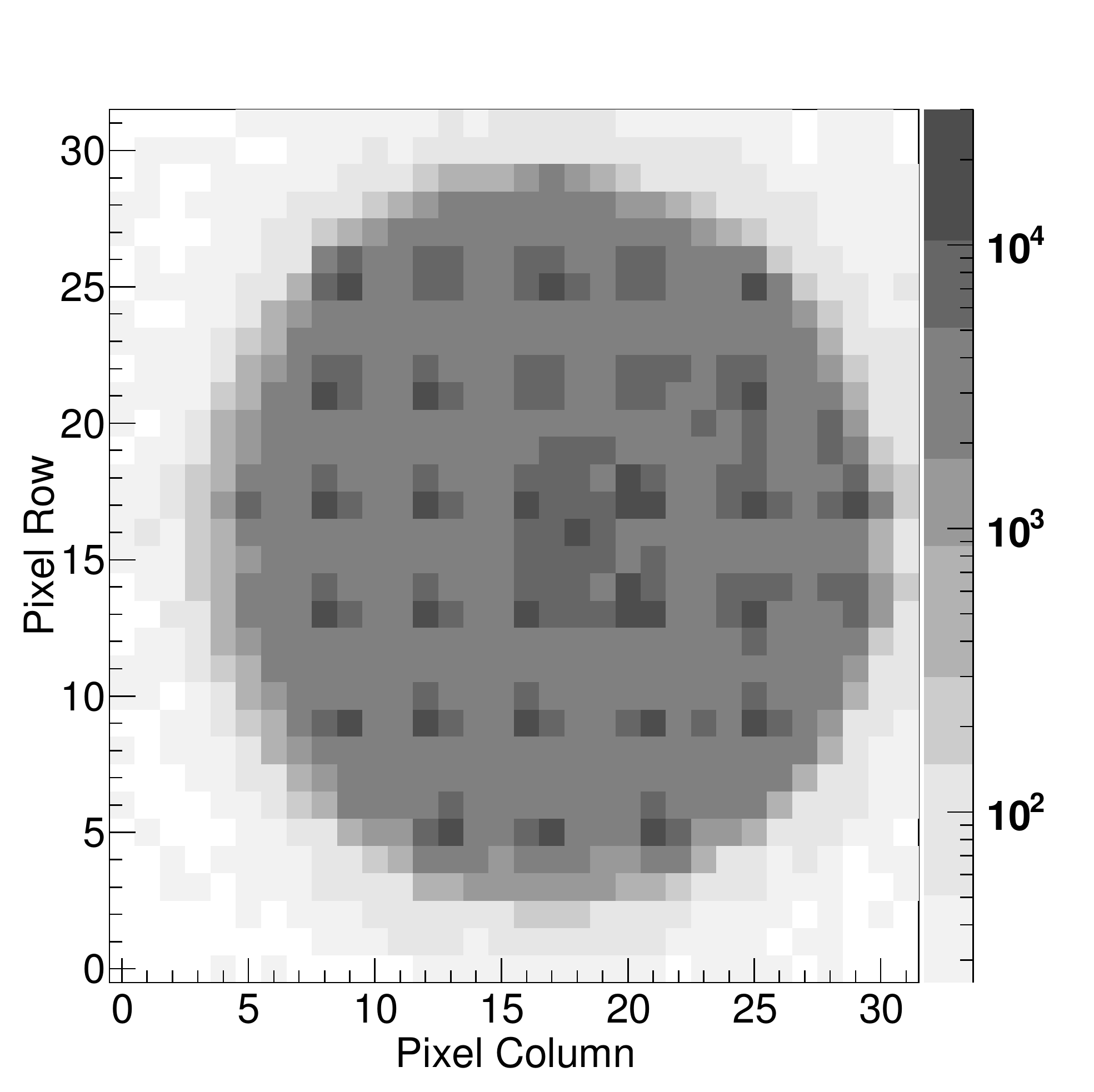} & 
\includegraphics [scale=0.35] {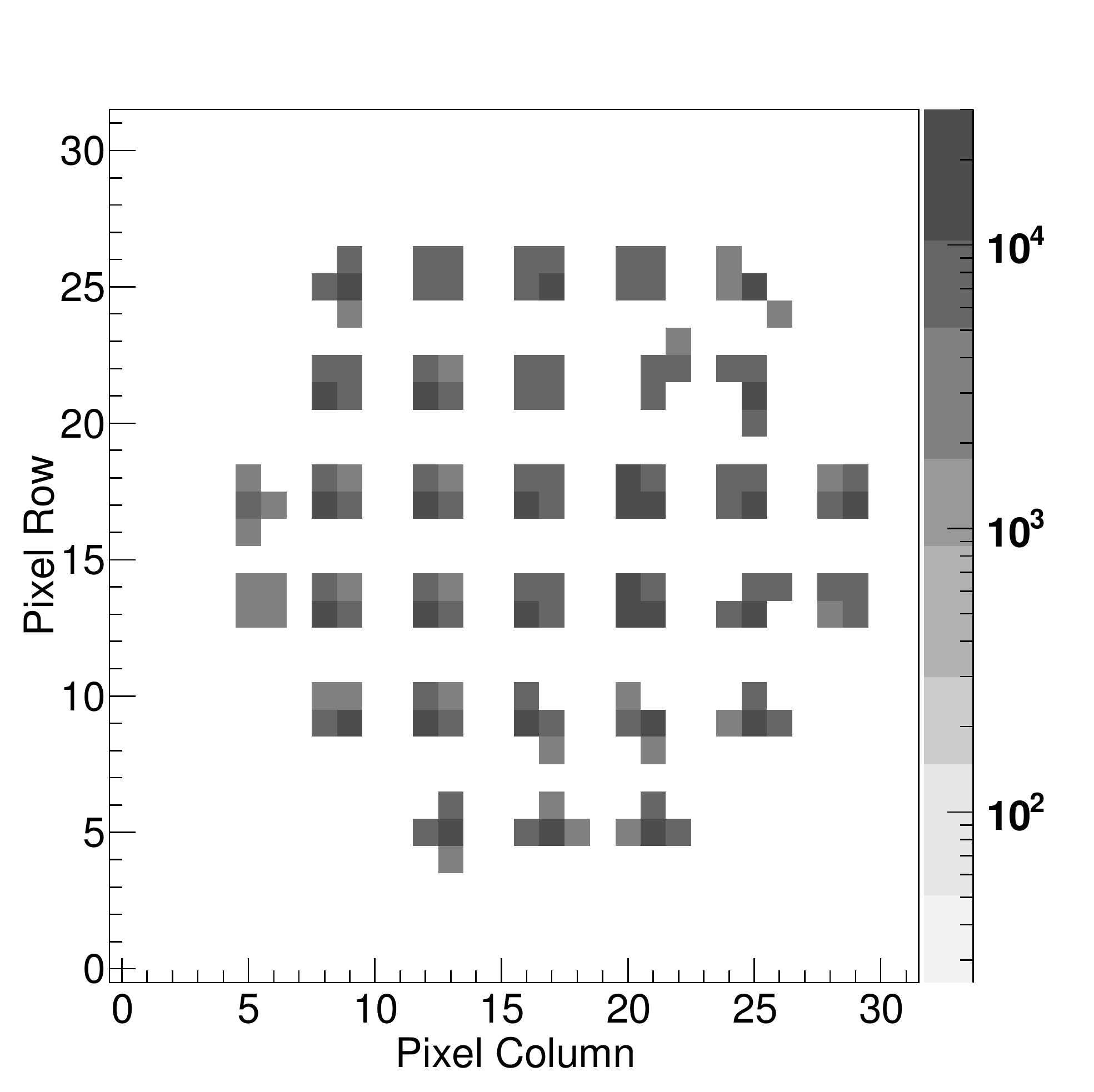}
\\
\end{tabular}
\caption{
Hit Map for one single HPD before (top figures) and after the
ion-feedback cuts (bottom figures). In the two top figures the
algorithm is not able to reconstruct clusters due to the
ion-feedback effect. After applying the cuts the ion-feedback is reduced and the
light spots are reconstructed correctly as shown in the bottom figures.} 
\label{fig:singleifb}
\end{figure}

A study to reduce this background has been performed analysing the behaviour of
two variables: the number of hits per event and the maximum cluster size per
event.  Comparing HPDs with large ion-feedback to HPDs without it, one can
define selection cuts to reject ion-feedback events.  A cut on the
number of hits larger than three and on the cluster size larger than
two has been chosen (see
Figure~\ref{fig:ifb}). The performance of the algorithm is shown in Figure~\ref{fig:singleifb} for one
single HPD affected by ion-feedback. The drop of the signal efficiency due to
  these cuts is negligible $(<2\%)$.

\begin{figure}[ht]
\centering
\begin{tabular}{cc}
\includegraphics [scale=0.35] {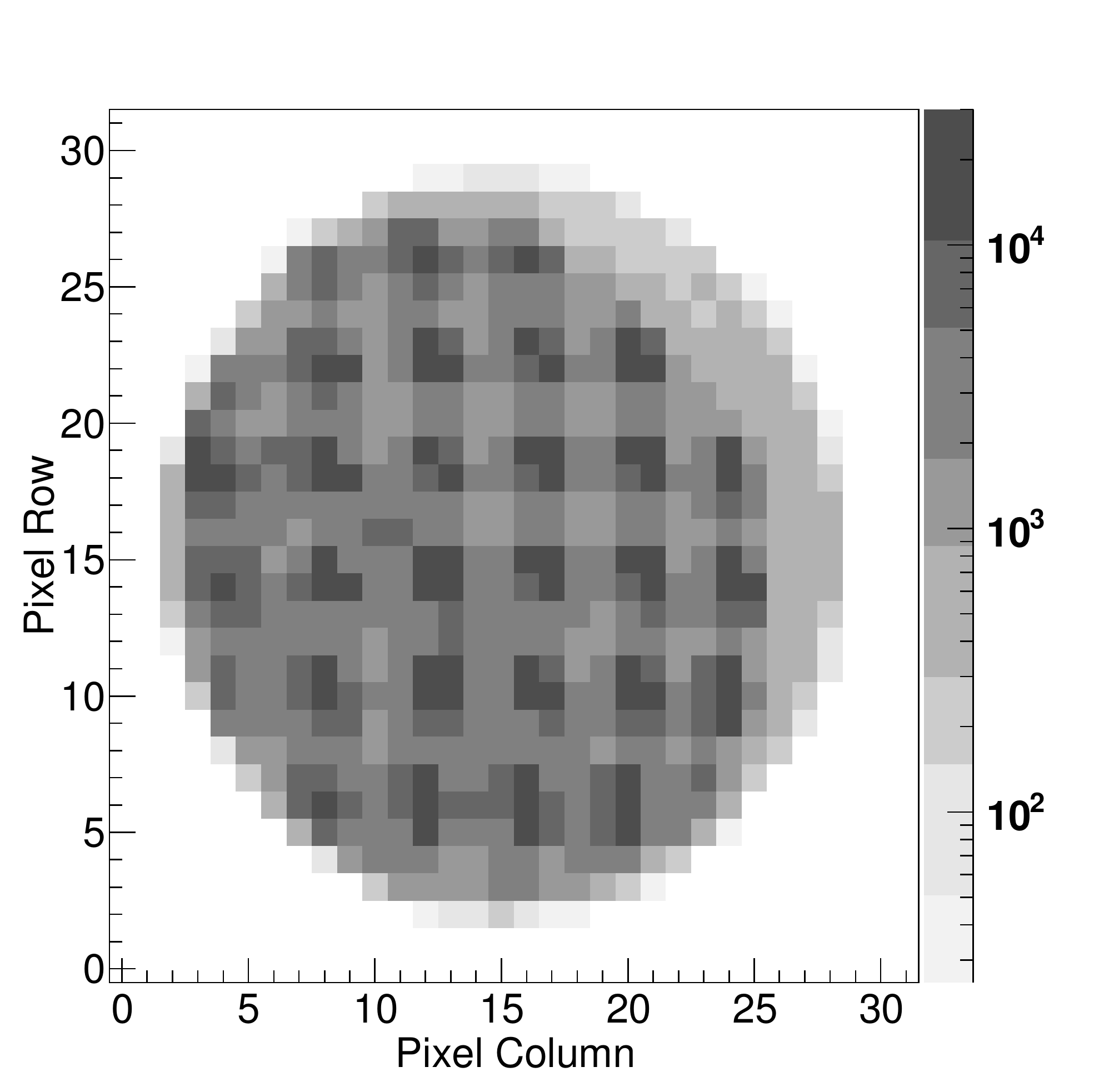} & 
\includegraphics [scale=0.35] {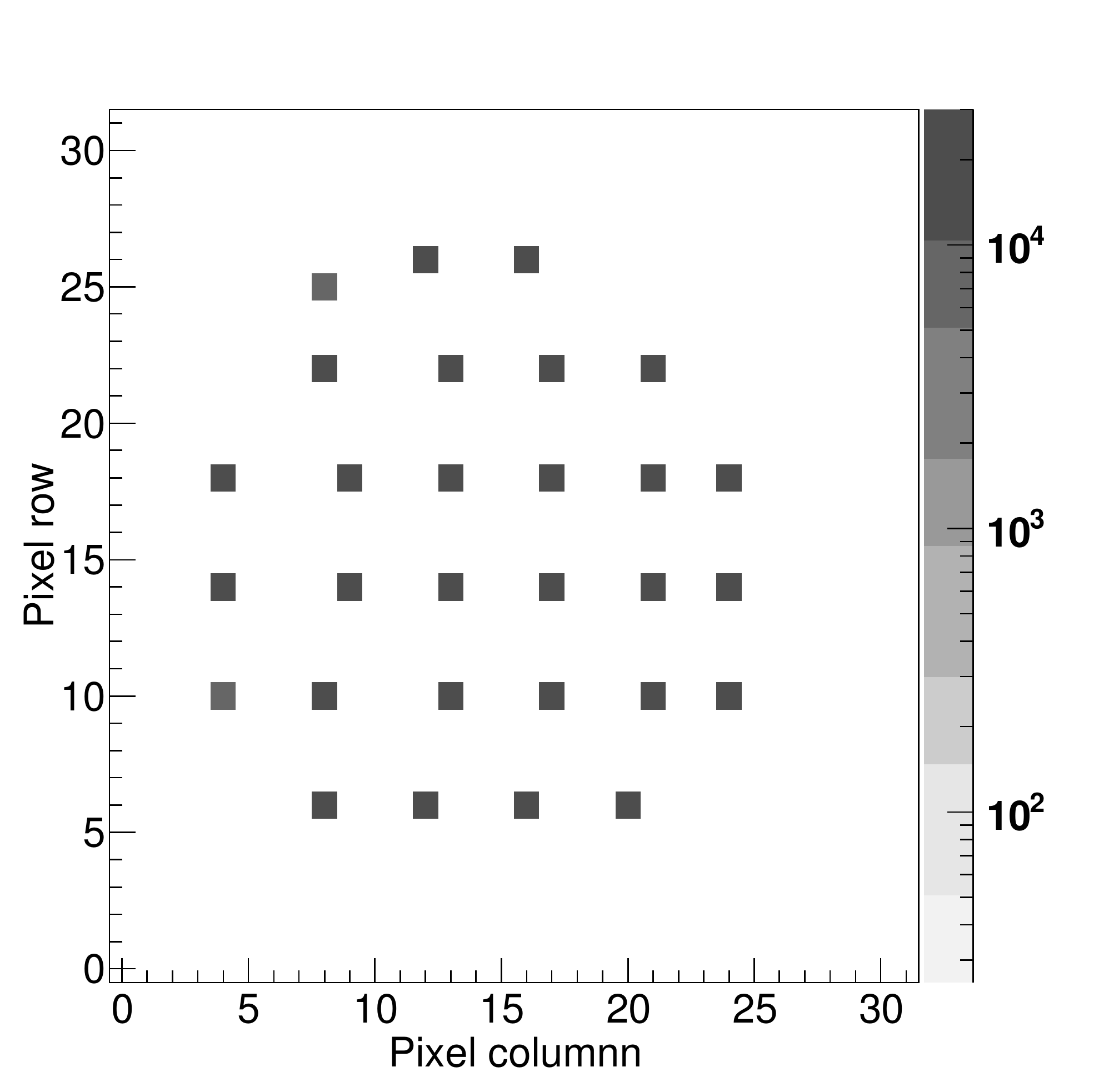} \\
\includegraphics [scale=0.35] {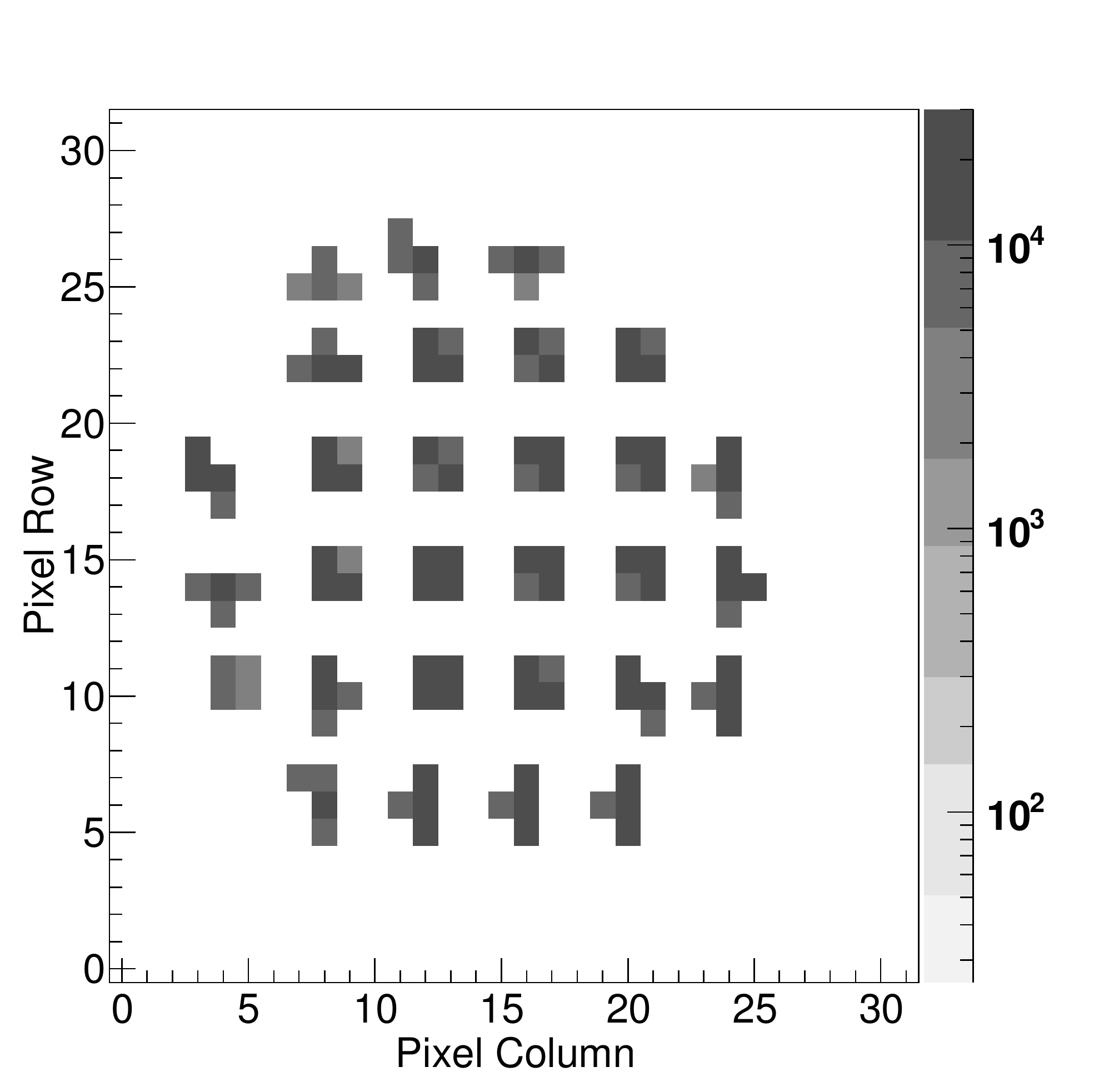} &
\includegraphics [scale=0.35] {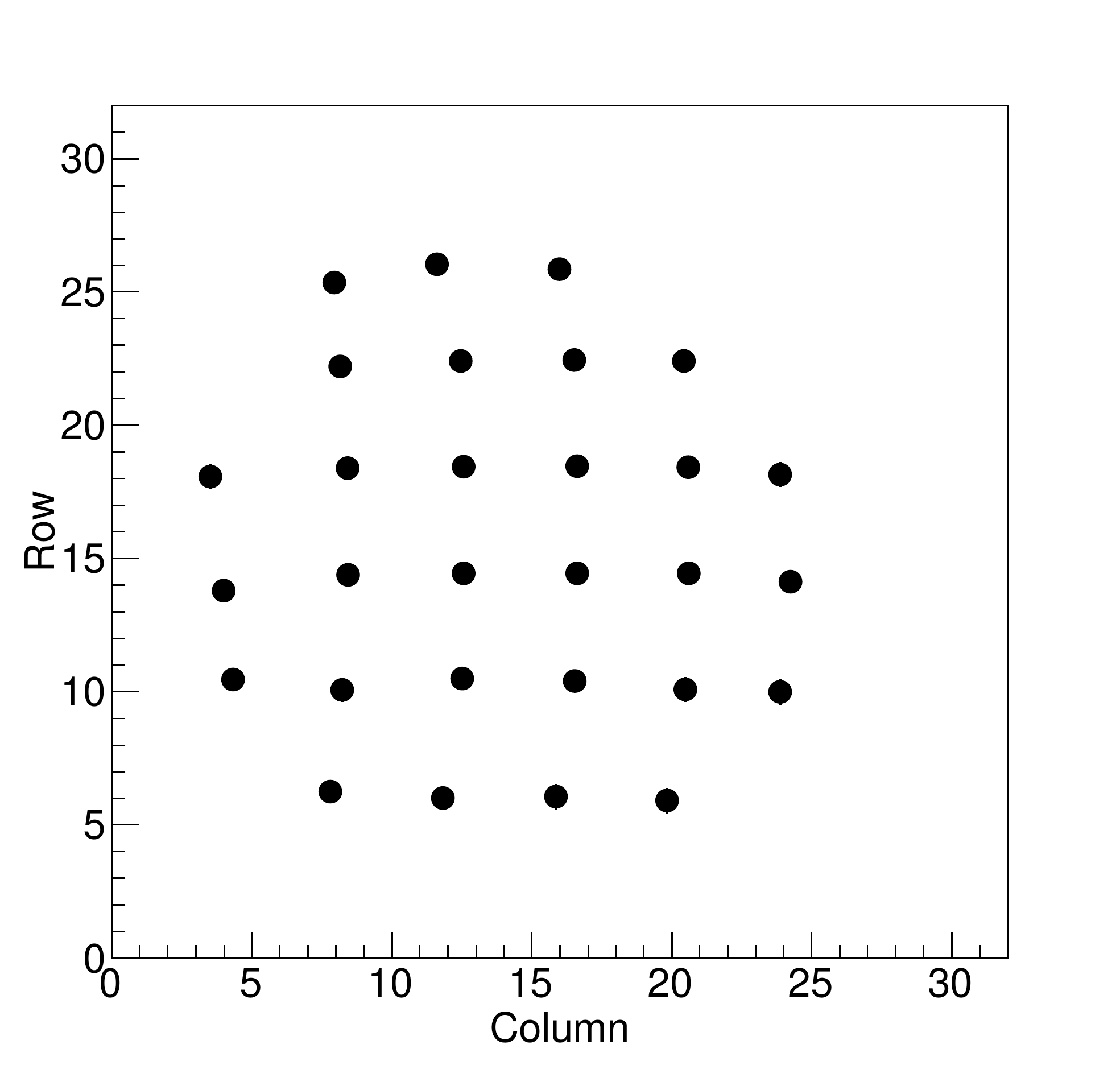} \\
\end{tabular}
\caption{
Light spot position reconstruction procedure: raw data (top left),
local reference maxima (top right),
clusters (bottom left) and cluster centres (bottom right).
} 
\label{fig:alghpd}
\end{figure}

After removing the noisy pixels and the effects of ion-feedback it is possible to
reconstruct the spots projected by the light projector as shown in
Figure~\ref{fig:alghpd}. 

\section{Correction of the Magnetic Distortions}
\label{sec:data}

Different data sets have been taken with the LHCb magnet on and off, at
full and half-field and with both magnet polarities.

\begin{figure}[htbp]
\centering
\begin{tabular}{cc}
\includegraphics [scale=0.37] {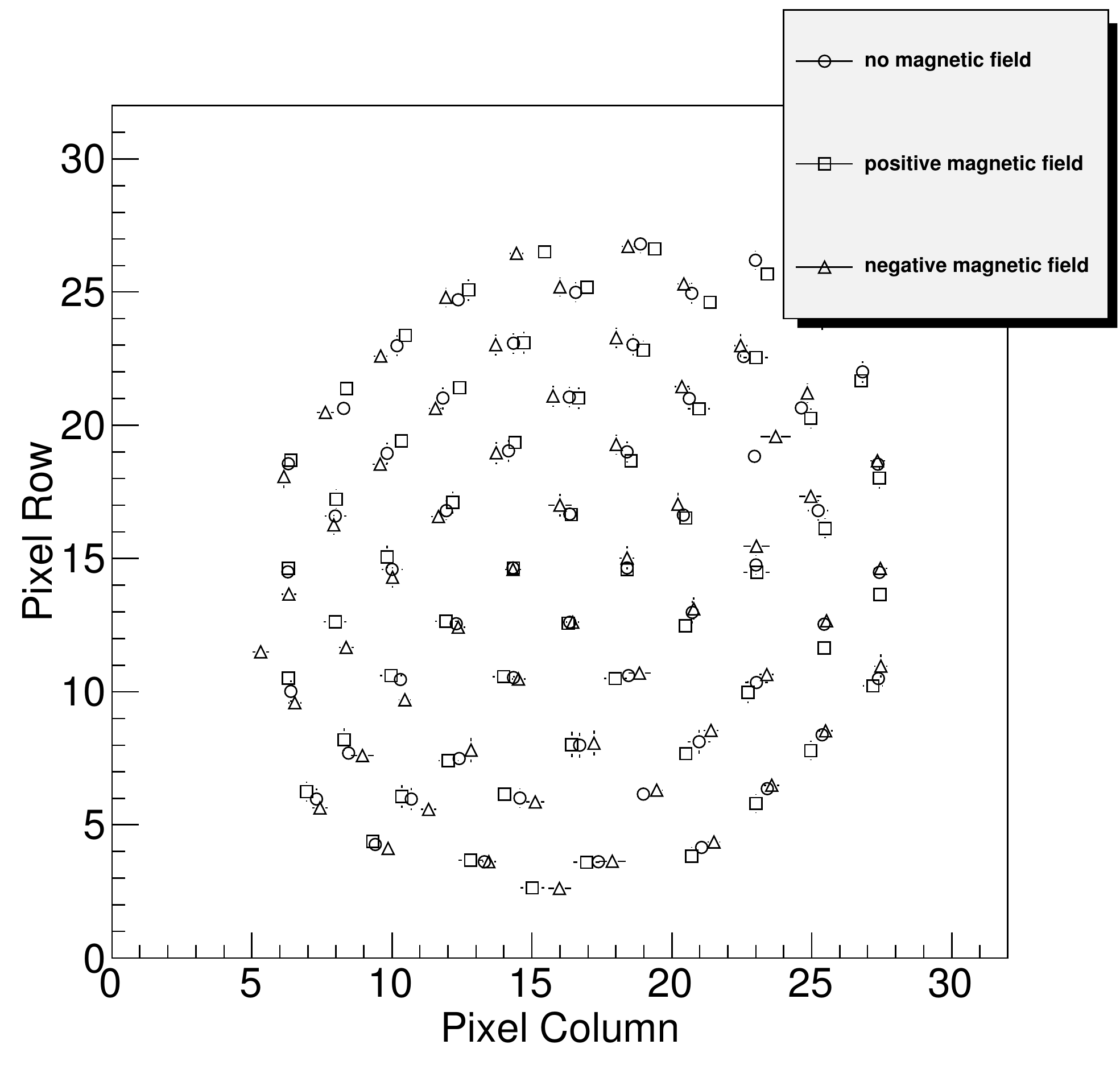} &
\includegraphics [scale=0.37] {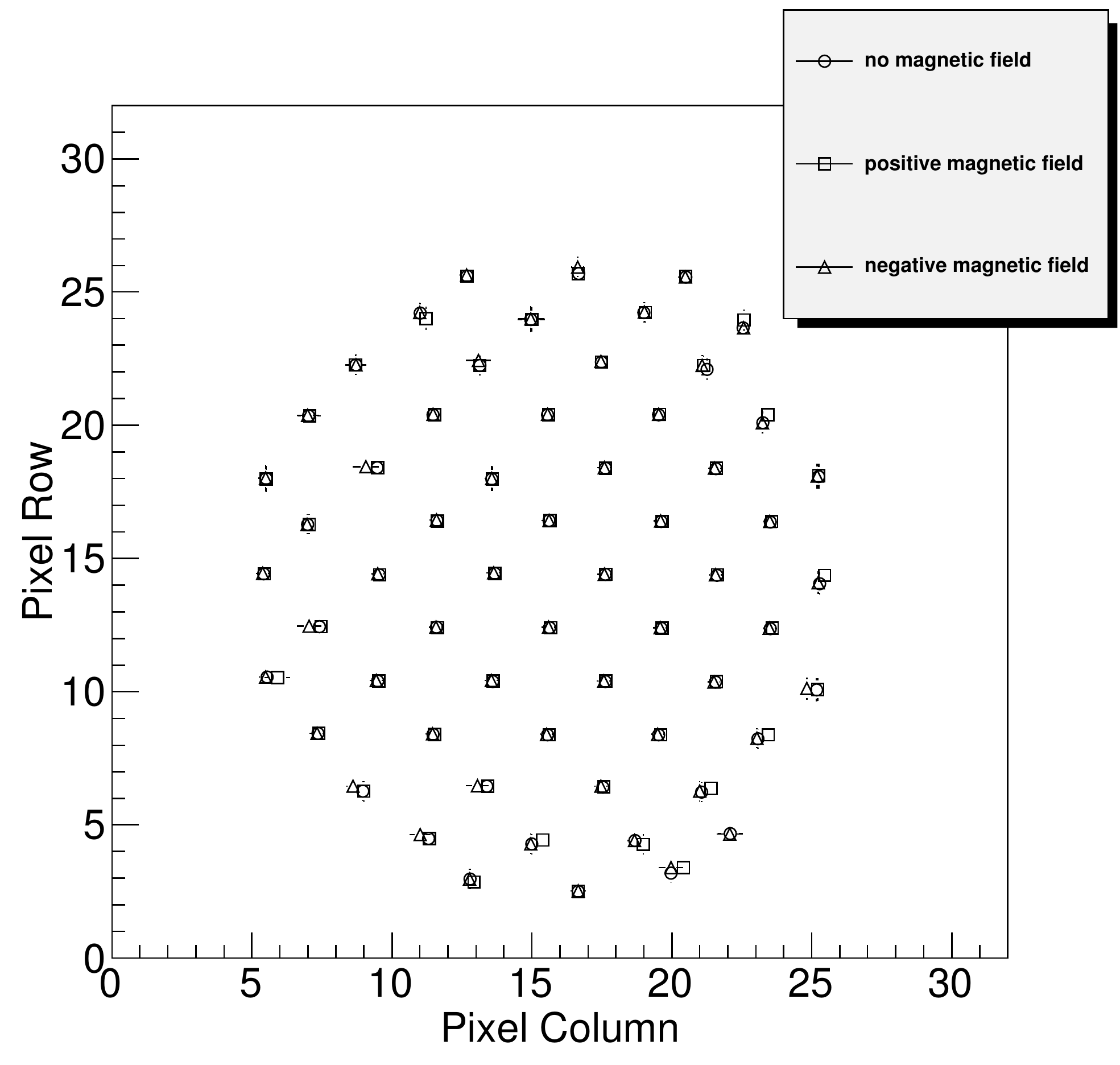} \\
\end{tabular}
\caption{
Reconstructed cluster centres for two different HPDs located in
two different positions inside the HPD matrix : 
square and triangle markers for magnetic field on (opposite polarities); 
circle marker for magnetic field off.} 
\label{fig:cgrot}
\end{figure}

A distortion effect is clearly visible in some of the HPDs.
As the stray magnetic field is not uniform in the HPD matrix,
the size of the effect
depends on the position of each tube inside the matrix. Two examples
are shown in Figure~\ref{fig:cgrot}.

\subsection{Parametrisation}

The expected image distortion in RICH2 is a non-uniform rotation and
stretch of the light spots with respect to
the HPD centre, while translations are expected to be negligible. This means that the
stray magnetic field has a non-negligible component along the HPD geometrical axis, while the
component perpendicular to it is small, in agreement with simulation
results~\cite{bi:TGnote}.

The distortion can be parametrised by a function of
the light spots position with respect to the HPD centre, expressed in polar coordinates $(r,\theta)$:
\beq
        \vec{d}_{i;j}(r,\theta) =
        {\Delta \theta}_{i;j}(r,\theta)\,\vec e_{\theta}+ {\Delta r}_{i;j}(r,\theta)\,\vec e_{r} 
\eeq
where $i$ denotes the HPD and $j$ denotes the light spot inside it. 

The average rotation angle for each HPD is given by
\beq
{\langle \Delta \theta \rangle}_{i} =
\frac{\sum_{j=0}^{n_{i}}{\Delta\theta}_{i;j}}{n_{i}} , 
\eeq 
where $n_{i}$ is the number of light spots reconstructed in the HPD $i$,
${\Delta\theta}_{i;j}$ is the rotation angle of the reconstructed light spot
$j$ with magnetic field on with respect to its position with magnetic field off
and calculated with respect to the centre of the 
HPD $i$.

\begin{figure}[htbp]
\centering
\begin{tabular}{c}
\includegraphics [scale=0.6] {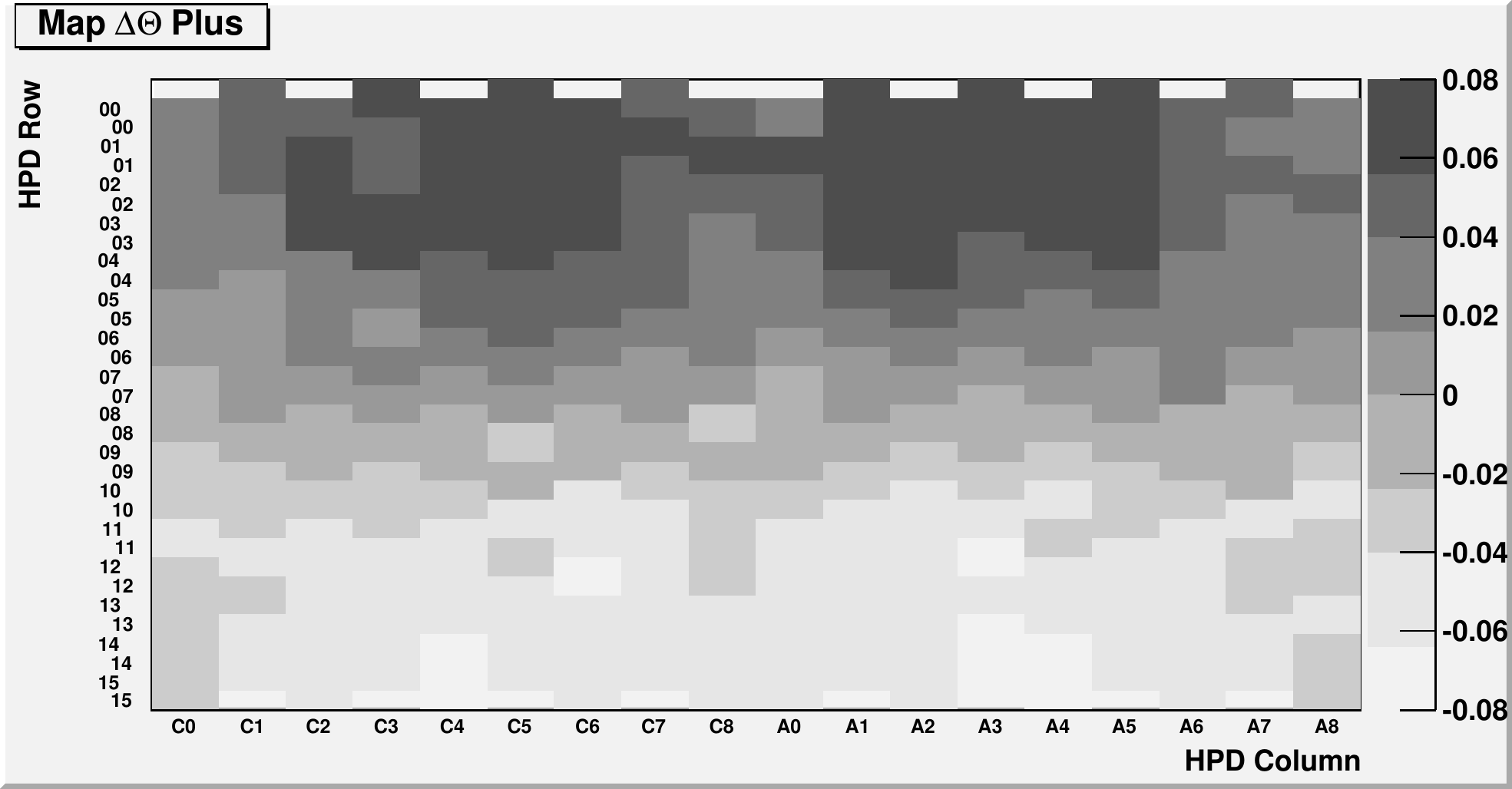}\\
\includegraphics [scale=0.6] {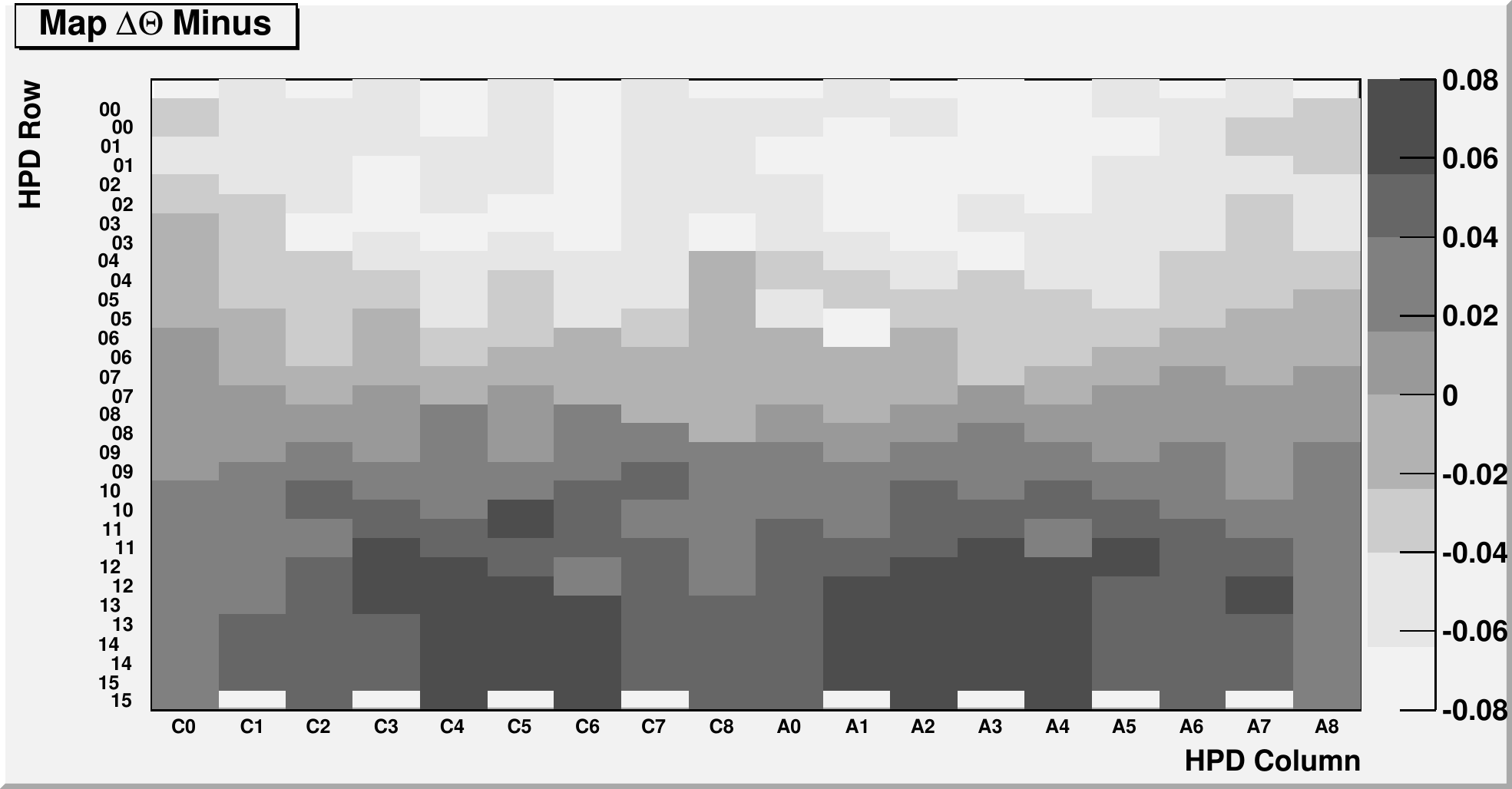}
\end{tabular}
\caption{Effective magnetic field map inside the HPD enclosure, determined from the average image rotation
angle for each HPD. 
Column and row numbers of the HPD position inside the
matrix are given. The HPDs in the matrix are positioned in an
hexagonal close pack arrangement. The double numbers in the vertical axis refer to
even and odd
columns that are vertically staggered.} 
\label{fig:maprot}
\end{figure}

The average rotation angle is related to the average magnetic field $
{\langle B \rangle}_{i} $ at the
location of the tube. If the magnetic field is small
and uniform inside the HPD volume, one can assume that the rotation angle is, to
a first approximation, proportional to the average magnetic field:
\beq\label{eq:avtheta} 
{\langle \Delta \theta \rangle}_{i}   \propto {\langle B \rangle}_{i} 
. 
\eeq
The sign of the rotation angle is related to the direction of the
magnetic field. 
The effective magnetic field map inside the HPD volume, estimated from
  the average image rotation angle, is shown in Figure~\ref{fig:maprot}.
The rotation angle changes its sign when going from the top to the bottom region
of the HPD matrix and it is close to zero in the central region.
The rotation effect
\mbox{$\langle \Delta \theta \rangle \lesssim 0.1\um{rad}$},
is clearly detectable and measurable.
Its value is not exactly symmetrical along the vertical direction since the
HPD matrix is not symmetrically located inside the overall shielding box.

\begin{figure}[htb]
\centering
\begin{tabular}{cc}
\includegraphics [scale=0.36] {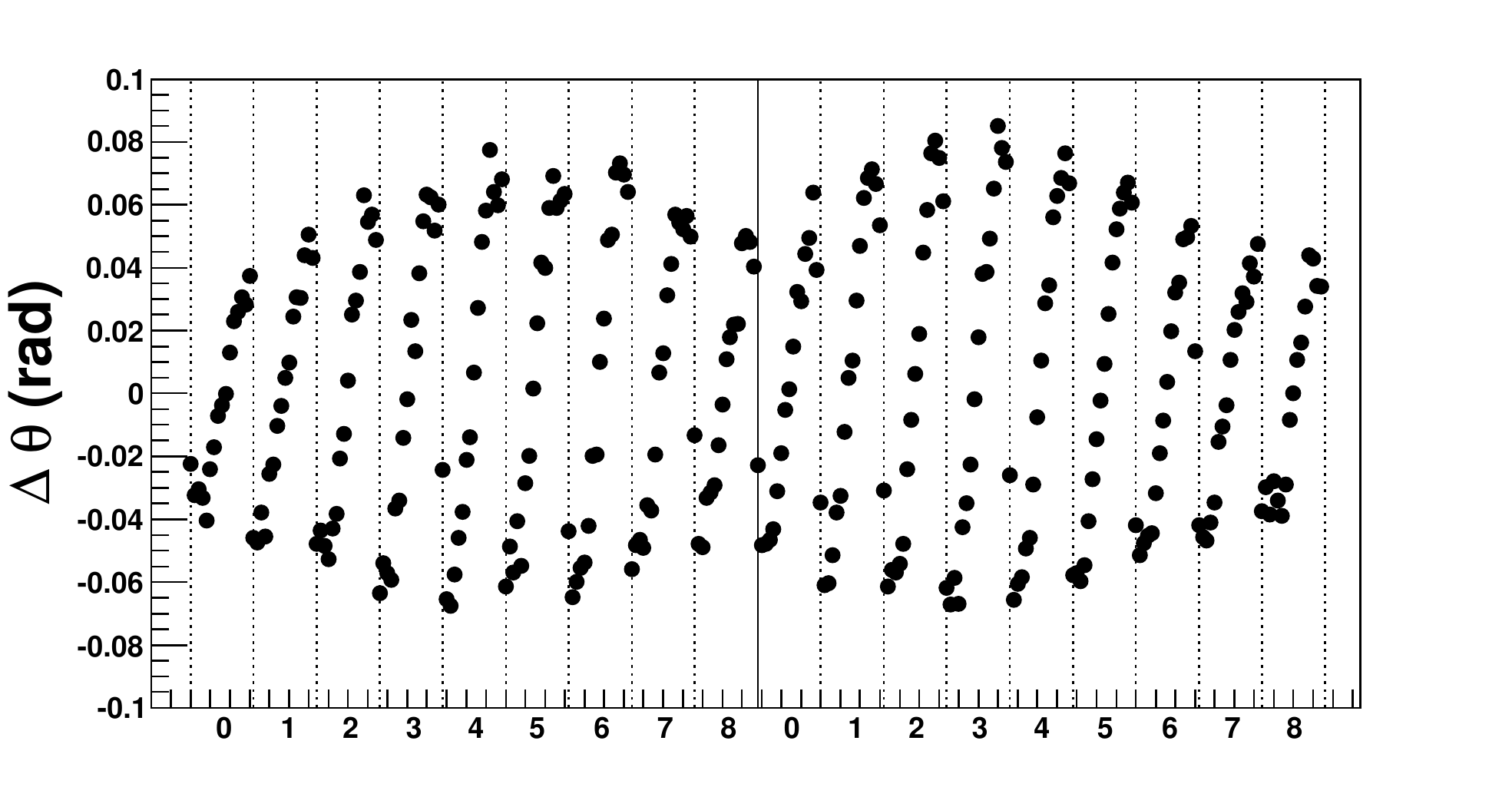}
\includegraphics [scale=0.36] {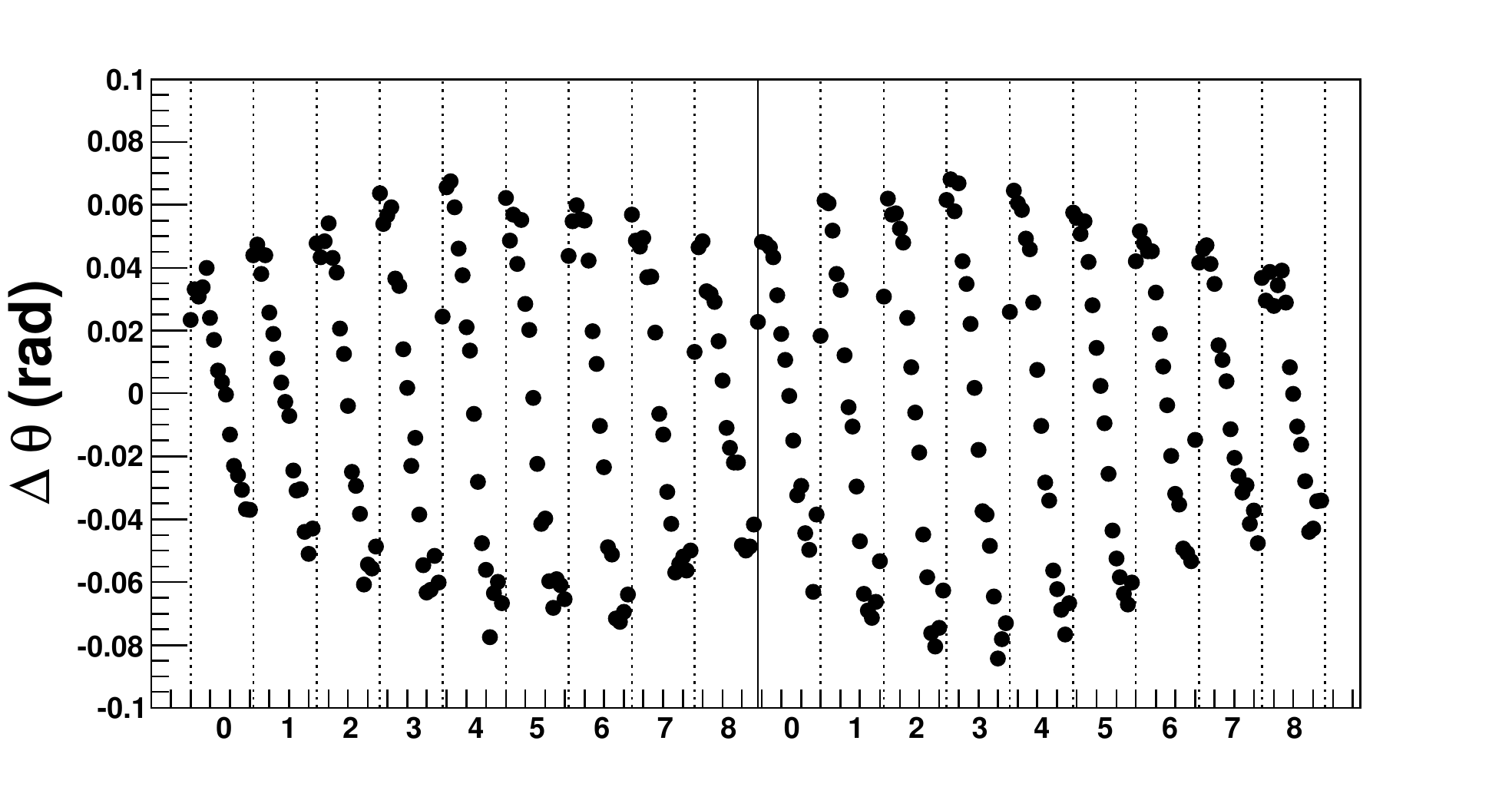}
\end{tabular}
\caption{HPD average image rotation angle for the two HPD matrices for positive
magnetic field (left) and negative magnetic field (right). The numbers on the
axis indicate the column number of the matrix.
The change of the magnetic field along each HPD column is clearly visible as well as
the fact that the magnetic field is stronger in the central region.} 
\label{fig:rotmedia}
\end{figure}

A modulation as a function of the HPD columns is also clearly visible (see Figure~\ref{fig:rotmedia}): the first
and the last column of each side show a smaller rotation angle with respect to
the central columns.  This is consistent with
the fact that the lateral columns, being closer to the vertical walls of the overall
shielding are affected
by a smaller magnetic field than the central columns.  The effect of the
overall shielding box is also evident for the HPDs installed on top or bottom of each
column.
All these effects are compatible with the results from simulations. 

Since the magnetic field intensity is different for each HPD, the distortion parameters are
different from HPD to HPD. However, after normalising the parameters to
the average rotation angle, which is proportional to 
the strength of the magnetic field, one
can find universal parameters for all HPDs.

\begin{figure}[htb]
\centering
\begin{tabular}{cc}
\includegraphics [scale=0.36] {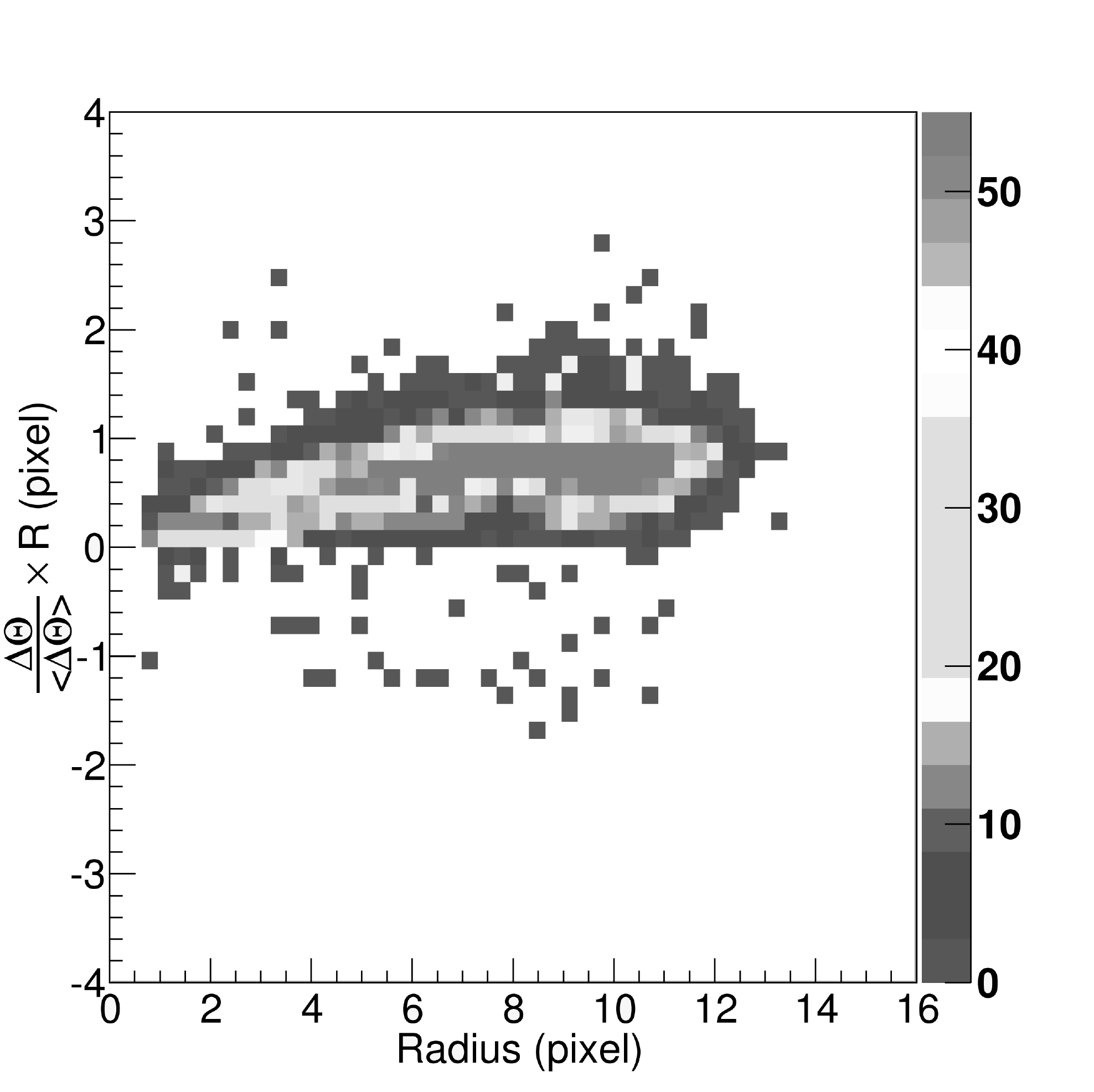}
\includegraphics [scale=0.36]
{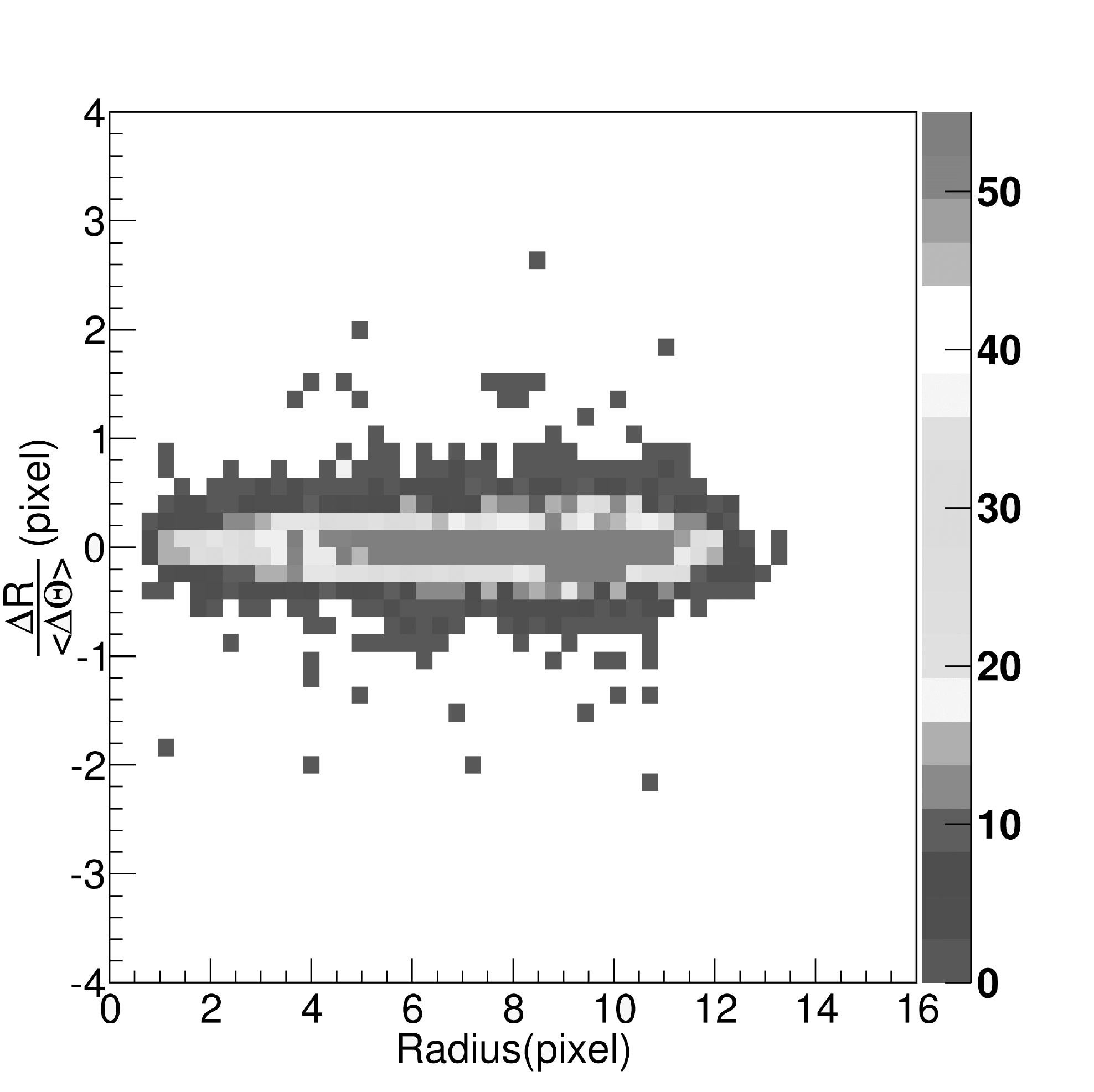}
\end{tabular}
\caption{
Rotation angle (left) and radius variation (right) of the light spots as a function of the distance from
the HPD centre.
} 
\label{fig:param}
\end{figure}

Two curves are needed to parametrise the distortion effect (see Figure~\ref{fig:param}): the
rotation angle and the radius variation of each single spot of the pattern:
\begin{equation}
\left\{
\begin{array}{ll}
    \Delta \theta_{i;j} & =  \langle \Delta\theta \rangle_{i}\, f_{\theta}(r_{i;j})\\
    \Delta r_{i;j} & = \langle \Delta\theta \rangle_{i}\,f_{r}(r_{i;j})
\end{array}
\right.
\end{equation}

In Figure~\ref{fig:param} the quantity $r_{i;j}\,{\Delta\theta}_{i;j}$ was used instead of
${\Delta\theta}_{i;j}$ 
since for small $r$ the error on the light spot angle is large, as distances
are measured, not angles.

From the first plot in Figure~\ref{fig:param} one can see that the quantity
$r\Delta \theta$ increases as a function of the radius
up to $r \sim 6 \um{pixel}$ and then it becomes approximately constant. 
From the second 
plot in Figure~\ref{fig:param} one can assume that, on average, there is no significant change
of the light spots distance from the centre. 

Therefore there is only a non-uniform rotation
and no stretch of the image within the current sensitivity.

In order to parametrise the angular distortion a fit with a second order
polynomial function has been performed: 
\beq 
f_\theta(r)=a\, r +b \, r^{2} 
\eeq 
obtaining: 
\beq 
a = 0.17 \pm 0.04 \um{pixel^{-1}}
\;\;\;\;
b=-0.009  \pm 0.004 \um{pixel^{-2}}
\eeq 

Therefore the magnetic distortion effects can be corrected applying to
all HPDs using the universal
relations

\begin{equation}
\left\{
\begin{array}{ll}
    \Delta \theta_{i;j} & =\langle \Delta\theta \rangle_{i}\, f_{\theta}(r_{i;j})/r_{i;j}
\\
        \Delta r_{i;j} & = 0
\end{array}
\right.
\end{equation}
with the same parameters regardless of their
position.
Moreover the same parameters are valid, with a sign inversion, for the two
opposite polarities of the magnetic field.

Measurements were also taken at half-nominal field values,
demonstrating that the rotation angle is proportional to the magnetic field intensity.

\subsection{Correction procedure}

Once the magnetic distortion effect has been parametrised, it is possible to
correct the data. The corrections have been
applied to the light spot positions with magnetic field on and then compared to the positions with
magnetic field off.
The inverse coordinate transformations from the non-corrected coordinates,
$ ( x_{0}^{(nc)} , y_{0}^{(nc)} ) $, to the corrected ones,
$ ( x^{(c)} , y^{(c)} ) $, via $ \Delta \theta_{i;j} $, are: 

\begin{equation}
\left\{
\begin{array}{ll}
     x^{(c)} = \,x_{0}^{(nc)} \cos(- \Delta \theta ) - \,y_{0}^{(nc)} \sin(- \Delta \theta )
\\
        y^{(c)} = \,x_{0}^{(nc)} \sin(- \Delta \theta ) + \,y_{0}^{(nc)} \cos(- \Delta \theta )
\end{array}
\right.
\end{equation}
where the indices $i$ and $j$ have been suppressed for clarity.

\begin{figure}[htb]
\centering
\begin{tabular}{cccc}
\includegraphics [scale=0.21] {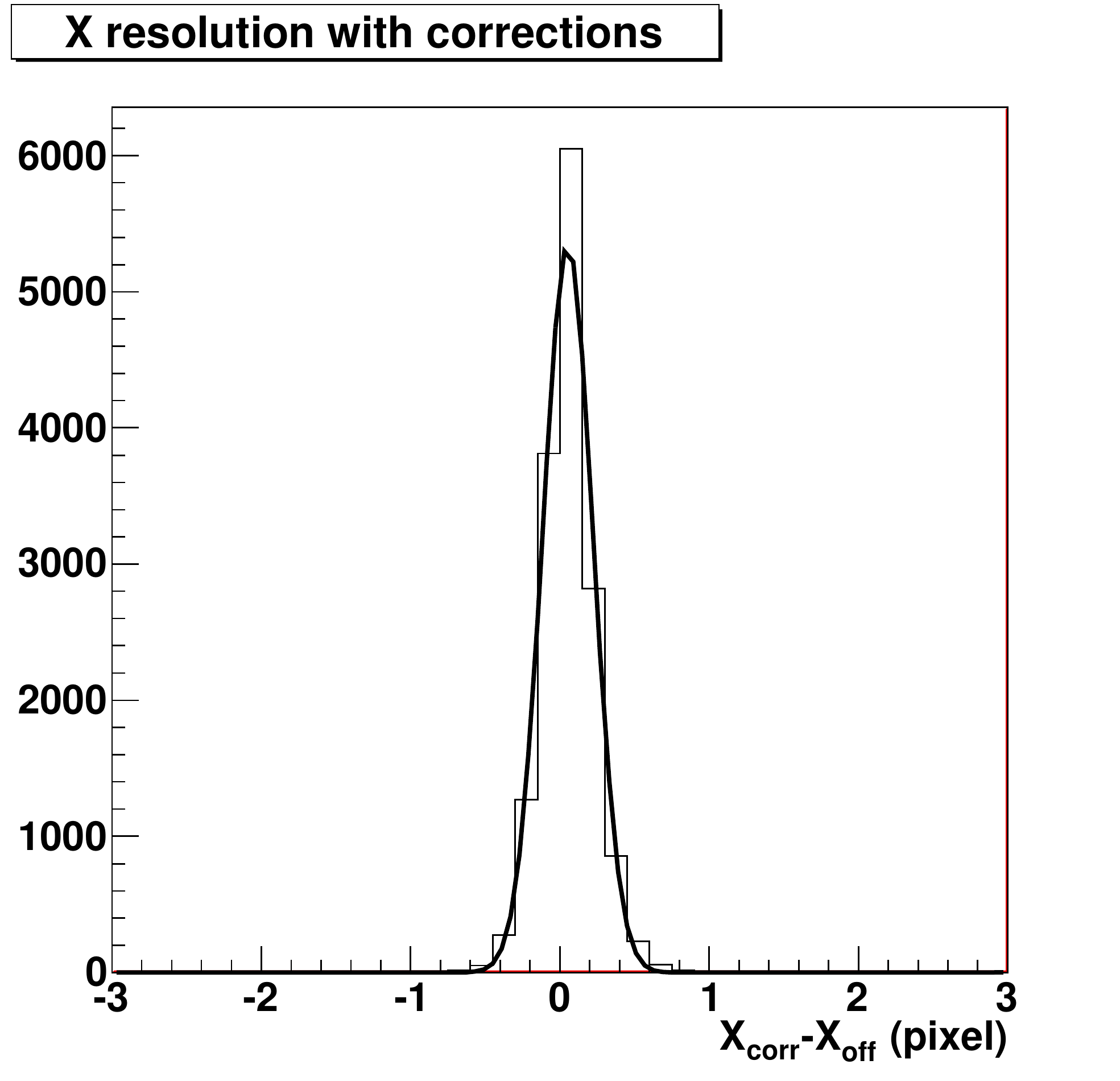}
&
\includegraphics [scale=0.21] {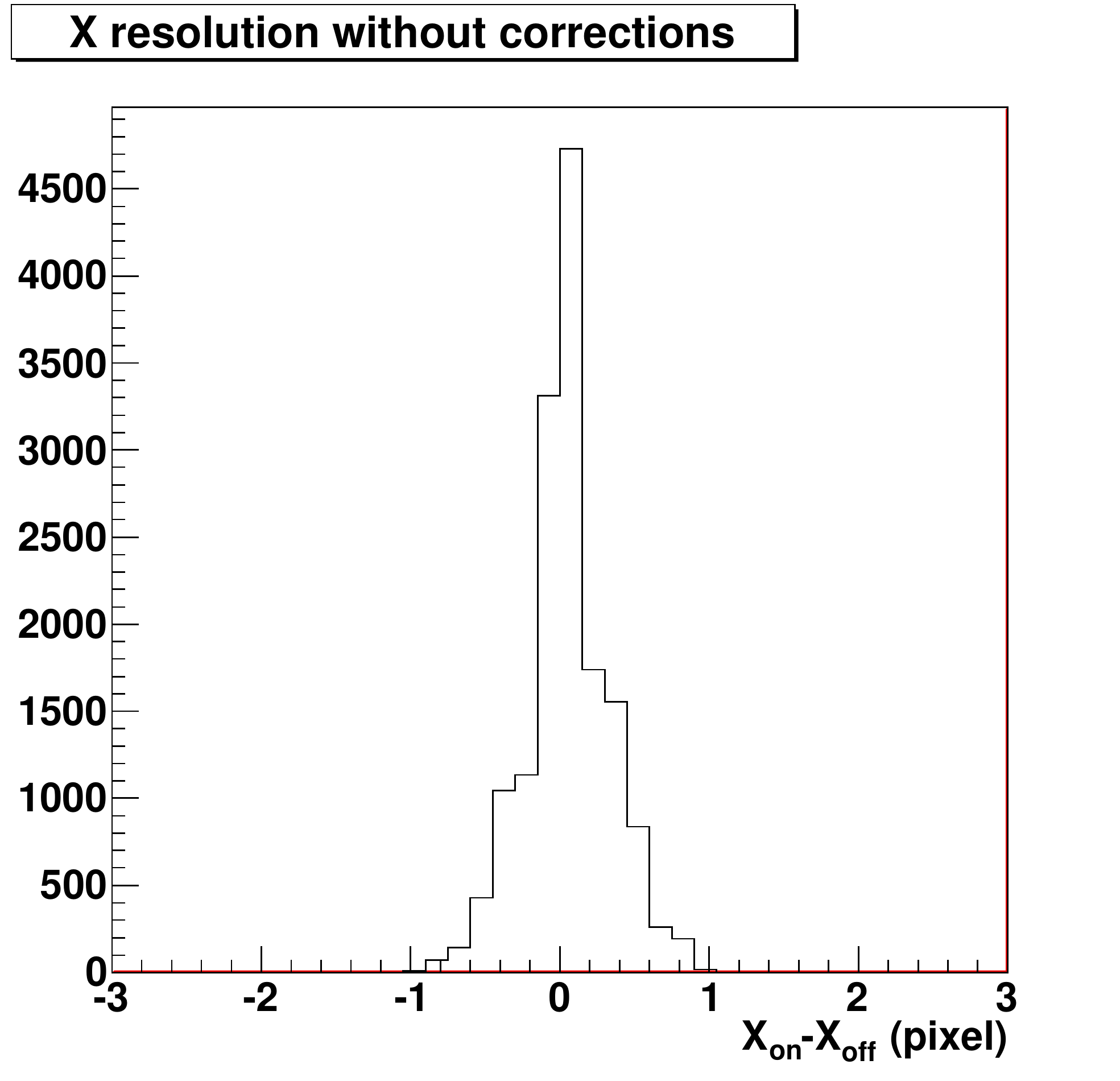}\\
\includegraphics [scale=0.21] {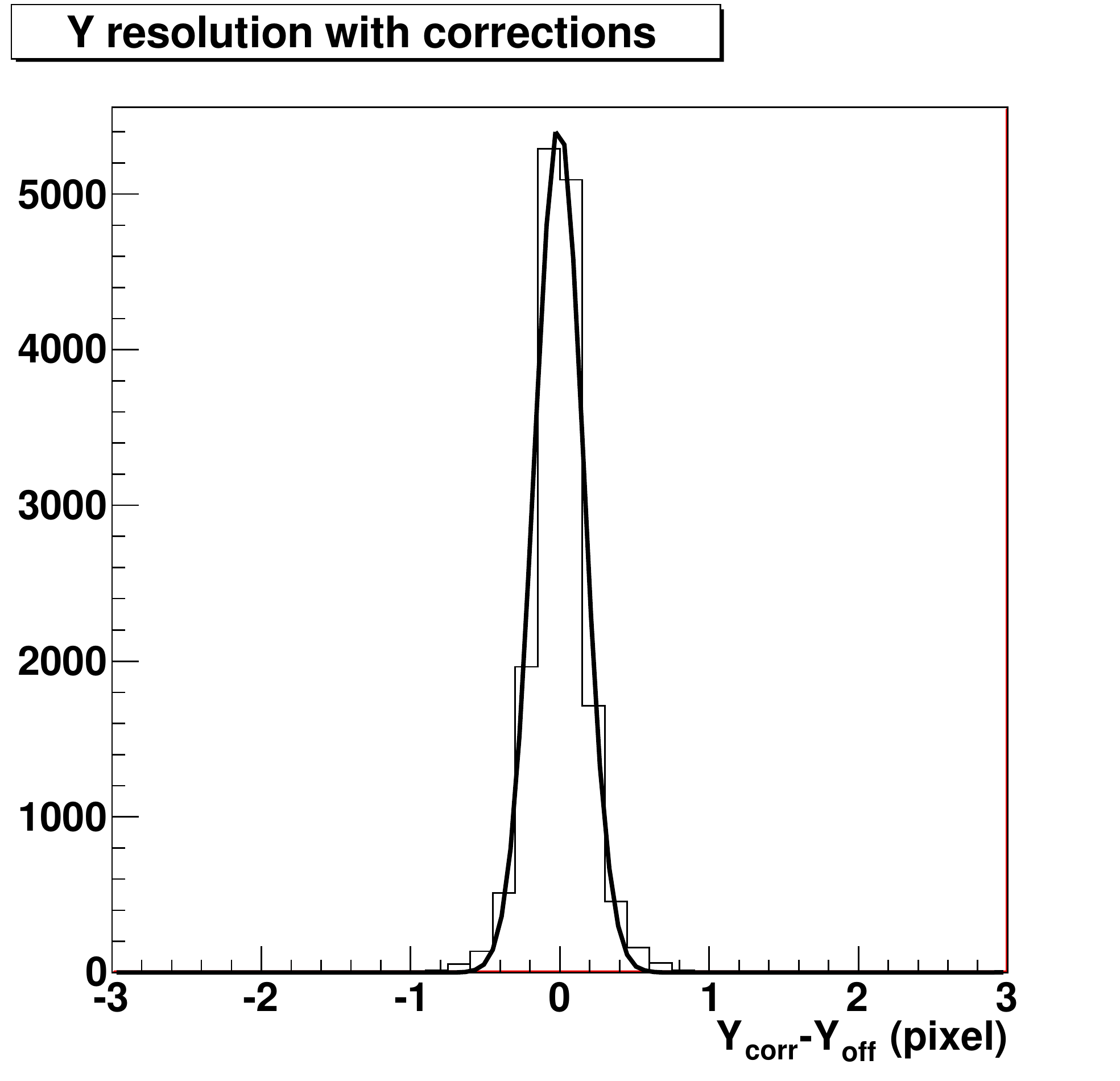}
&
\includegraphics [scale=0.21] {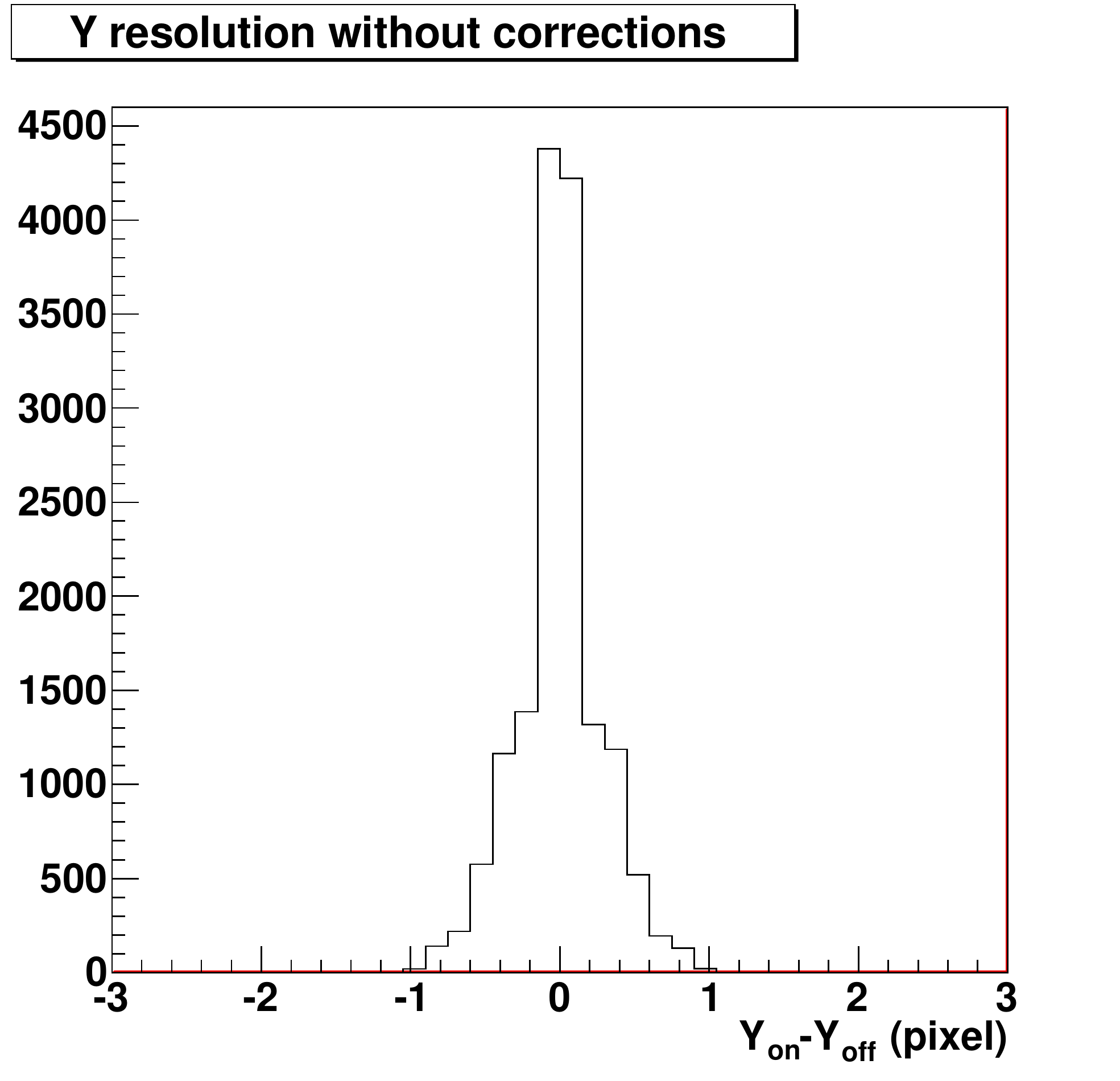}
\\
\end{tabular}
\caption{
Resolution, in the $x$ and $y$ coordinates, with (left) and without
(right) applying the correction
procedure (in pixels).
}
\label{fig:corr}
\end{figure}

The resolution obtained after the correction procedure (see the two
left plots of Figure~\ref{fig:corr}), amounts to 
\beq
\sigma_{x} \simeq 0.18 \um{pixel}
\;\;\;\; 
\sigma_{y}\simeq 0.18 \um{pixel}
\eeq

This resolution must be compared to the one found
without any correction of the magnetic field effects, (see the two right plots of
Figure~\ref{fig:corr}), that is:
\beq 
        \mathrm{RMS}_{x}\simeq \mathrm{RMS}_{y} \simeq 0.33 \um{pixel} 
\eeq 
For the sake of comparison the resolution due to the finite pixel size
is $\sigma \simeq 0.29 \um{pixel}$.
The correction procedure therefore restores the optimal resolution of
the Cherenkov angle~\cite{bi:Thesis}.
\section{Correction of the magnetic distortion in the first LHCb data}

The correction parameters, determined as described before, are stored
in the LHCb condition database and were used on the first
data taken by LHCb in 2010.

Saturated Cherenkov rings were reconstructed in minimum bias data,
taken with magnetic field on and with both polarities.

The rings were reconstructed with and without applying the magnetic field
distortion correction.  The measured Cherenkov angle distribution of saturated
Cherenkov rings was fitted with a Gaussian plus a second-order polynomial. The
standard deviation of the Gaussian fit is summarised in
Table~\ref{ta:ResolData}. As the residual stray magnetic field is small in the
central region of the HPD matrix, the data in Table~\ref{ta:ResolData}
show that the correction has
little effect in that region. On the other hand the residual stray magnetic field is not
negligibly small in the external regions of the HPD matrix, where the correction improves the
resolution.

The correction parameters restore the optimal shape of the ring that otherwise would be
distorted. This decreases the standard deviation of the measured
Cherenkov angle distribution improving the capability of the PID system.
\begin{table}[hbt]
\centering
\caption{
Standard deviation of the Gaussian plus second order polynomial fit of saturated Cherenkov rings without and with magnetic distortion corrections.
}
\begin{tabular}{||c|c|c||}
\hline
\multicolumn{3}{|c|}{\textbf{Magnetic field down}} \\ 
\hline
                        & external         & central         \\
\hline
Without correction      & 0.81\um{mrad} & 0.73\um{mrad}  \\
With correction         & 0.76\um{mrad} & 0.73\um{mrad}  \\
\hline
\multicolumn{3}{|c|}{\textbf{Magnetic field up}} \\ 
\hline
                        & external         & central         \\
\hline
Without correction      & 0.76\um{mrad} & 0.70\um{mrad}  \\
With correction         & 0.73\um{mrad} & 0.70\um{mrad}  \\
\hline
\end{tabular}

\label{ta:ResolData}
\end{table}

\section{Conclusions}

A system to measure the effects of the magnetic distortions in the
RICH2 detector of LHCb has been set up.  

The residual stray magnetic field, that has a predominant component parallel to the HPD geometrical axis,
causes a small rotation of the photoelectrons from the nominal trajectory given
by the HPD electrostatic field.

By applying the
correction procedure described in this note the resolution improves from $\sim 0.33\um{pixel}$ to $\sim
0.18\um{pixel}$.

The effect of the correction has been cross-checked on real data. It improves the Cherenkov
angle resolution especially in the external region of the HPD
matrix. The optimal particle identification capability of RICH2 is restored.

It is planned to repeat the projector measurements during each shut-down periods
to check the stability of the parameters and look for possible hysteresis effects.

\section{Acknowledgements}

We thank the technical and administrative staff at
CERN and at the LHCb institutes, and acknowledge support from the
National Agencies, in particular 
CERN and
INFN (Italy).
We also thank C. Matteuzzi and O. Ullaland for many useful discussions
and support.








\end{document}